\documentclass[a4paper,11pt]{article}
\pdfoutput=1

\usepackage[a4paper,left=2.73cm,right=2.7cm,top=3cm,bottom=3.5cm]{geometry}

\usepackage{amsmath,amssymb,graphicx,wrapfig,caption,subcaption,enumerate,color,tikz}
\usetikzlibrary{arrows}
\usepackage{colortbl}
\usepackage{booktabs}
\usepackage{cite}
\usepackage{wrapfig}
\usepackage[mathscr]{euscript}

\usepackage[colorlinks=true,linktocpage=true,linkcolor=blue,citecolor=blue]{hyperref}

\definecolor{gray}{rgb}{.9,.9,.9}
\tikzstyle{block} = [rectangle, text width=5em, text  centered, rounded corners]
\tikzstyle{line} = [draw, very thick, -latex']
\tikzstyle{RGflow}= [->, shorten <=1pt, thick, dashed, color=black!70]

\addtocontents{toc}{\protect\setcounter{tocdepth}{2}}
\numberwithin{equation}{section}

\newcommand{\bea}{\begin{eqnarray}}
\newcommand{\beal}[1]{\begin{eqnarray}\label{#1}}
\newcommand{\eea}{\end{eqnarray}}
\newcommand{\be}{\begin{equation}}
\newcommand{\bel}[1]{\begin{equation}\label{#1}}
\newcommand{\ee}{\end{equation}}

\newcommand{\bit}{\begin{itemize}}
\newcommand{\eit}{\end{itemize}}
\newcommand{\ben}{\begin{enumerate}}
\newcommand{\een}{\end{enumerate}}

\newcommand{\mt}[1]{\textrm{\tiny #1}}
\newcommand{\nc}{N_\mt{c}}

\newcommand{\sac}{\, , \qquad}
\newcommand{\eqn}[1]{Eq.~(\ref{#1})}

\newcommand{\fig}[1]{Fig.~\ref{#1}}

\newcommand{\bal}{\begin{align}}
\newcommand{\eal}{\end{align}}
\newcommand{\bse}{\begin{subequations}}
\newcommand{\ese}{\end{subequations}}

\def\Pt{{{P}}_T}
\def\Pl{{{P}}_L}

\begin{document}

\begin{titlepage}

\thispagestyle{empty}

\begin{flushright}
\end{flushright}

\vspace{40pt}  
	 
\begin{center}

{\LARGE \textbf{Domain Collisions}}
	\vspace{30pt}
		
{\large \bf  
Yago Bea,$^{1}$ Jorge Casalderrey-Solana,$^{2}$ Thanasis Giannakopoulos,$^{3}$ \\ David Mateos,$^{2,\,4}$  Mikel Sanchez-Garitaonandia,$^{2}$ and Miguel Zilh\~ao$^{3}$}

\vspace{25pt}

{ $^{1}$ Department of Physics and Helsinki Institute of Physics,
PL 64, FI-00014 University of Helsinki, Finland.}\\ 
\vspace{15pt}
{$^{2}$ Departament de F\'\i sica Qu\'antica i Astrof\'\i sica and Institut de Ci\`encies del Cosmos (ICC),\\  Universitat de Barcelona, Mart\'\i\  i Franqu\`es 1, ES-08028, Barcelona, Spain.}\\
\vspace{15pt}
{ $^{3}$ Centro de Astrof\'{\i}sica e Gravita\c c\~ao (CENTRA), \\
  Departamento de F\'{\i}sica, Instituto Superior T\'ecnico,
  Universidade de Lisboa, \\
  Av.\ Rovisco Pais 1, 1049-001 Lisboa,
  Portugal.}\\
\vspace{15pt}
{ $^{4}$ Instituci\'o Catalana de Recerca i Estudis Avan\c cats (ICREA), \\ Passeig Llu\'\i s Companys 23, ES-08010, Barcelona, Spain.}\\
\vspace{15pt}

\vspace{40pt}
				
\abstract{
We use holography to study collisions of phase domains formed in a four-dimensional, strongly-coupled gauge theory with a first-order, thermal phase transition. We find three qualitatively different dynamical regimes depending on the collision velocity. For low velocities the domains  slow down before the collision and subsequently merge and relax to equilibrium. For intermediate velocities no slow down is present before the merger. For high enough velocities the domains can collide and break apart several times before they finally merge. These features leave an imprint on the time evolution of the entropy of the system, which we compute from the area of the dual horizon on the gravity side.
}

\end{center}

\end{titlepage}

\tableofcontents

\hrulefill
\vspace{10pt}

\section{Introduction}
\label{intro}
\begin{figure}[t]
\begin{center}
	\includegraphics[width=0.7\textwidth]{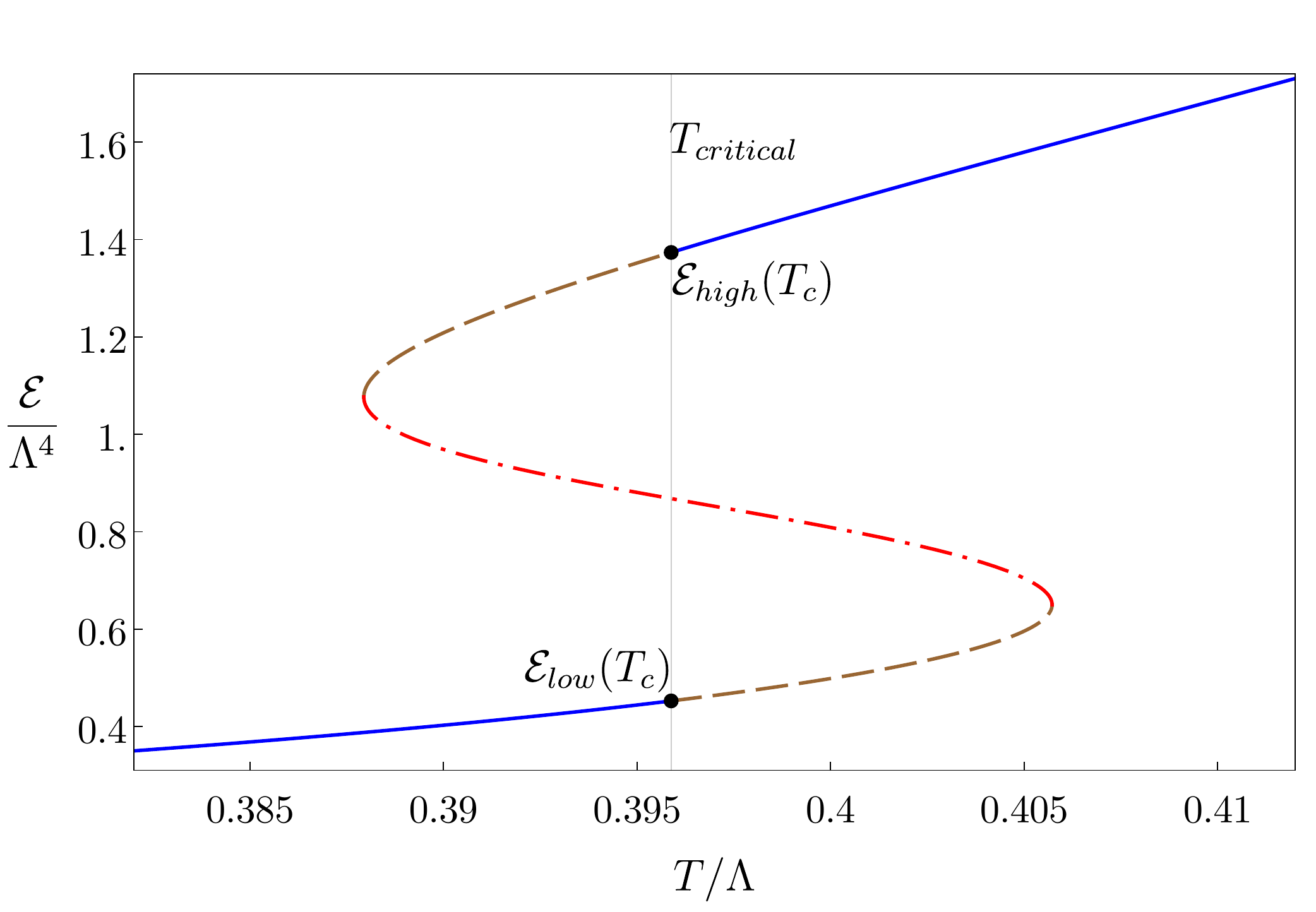} 
\end{center}
\vspace{-5mm}
\caption{\label{fig:EvsT} \small Energy as a function of  temperature for the theory considered in this paper, with intrinsic scale $\Lambda$. Stable, metastable and unstable states are represented by solid, dashed and dot-dashed curves, respectively. The vertical grey line corresponds to the phase transition temperature.}
\vspace{2mm}
\end{figure}

\begin{figure}[h!!!]
\begin{center}
\includegraphics[width=.7\textwidth]{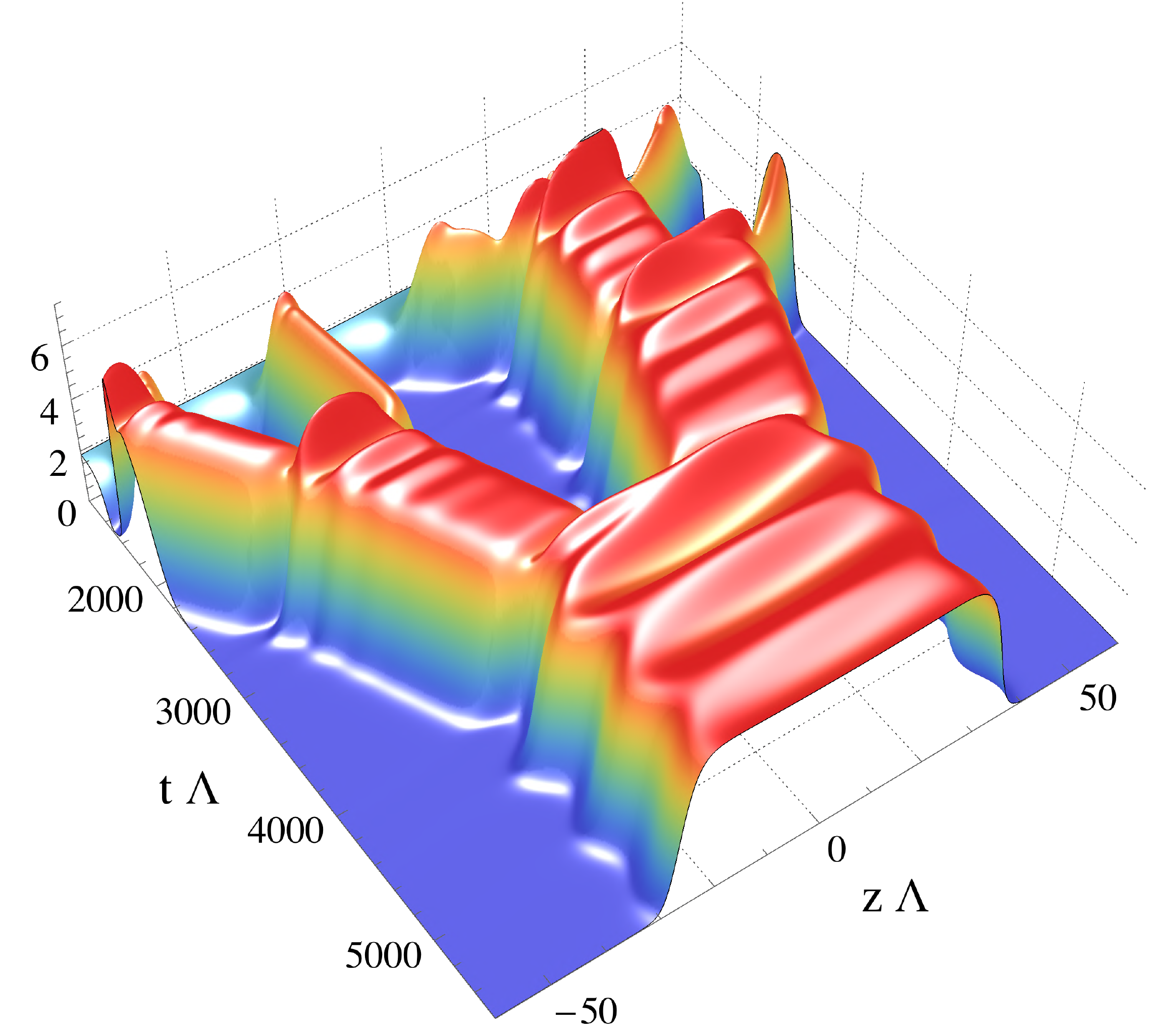} 
\put(-320,210){\mbox{{$10^2$ $\mathcal{E}/\Lambda^4$}}}
\end{center}
\vspace{-5mm}
\caption{\label{reflecting1} 
\small Evolution of the energy density as a function of time and space for an initial homogeneous state in the spinodal region that is slightly perturbed at $t=0$.  Figure taken from \cite{Attems:2019yqn}.
}
\vspace{7mm}
\end{figure}

Holography has provided numerous insights into the out-of-equilibrium properties of hot, strongly-coupled, non-Abelian plasmas \cite{Chesler:2008hg,Chesler:2009cy,Heller:2011ju,Chesler:2010bi,Heller:2012km,Heller:2013oxa,Casalderrey-Solana:2013aba,Casalderrey-Solana:2013sxa,Chesler:2015wra,Chesler:2015bba,Chesler:2015lsa,Buchel:2015saa,Chesler:2016ceu,Attems:2016tby,Casalderrey-Solana:2016xfq,Attems:2016ugt,Attems:2017zam,Attems:2017ezz,Janik:2017ykj,Rougemont:2017tlu,Gursoy:2016ggq,Critelli:2018osu,Buchel:2018ttd,Czajka:2018bod,Czajka:2018egm,Attems:2018gou,Attems:2019yqn,Bantilan:2020pay,Bea:2020ees,Casalderrey-Solana:2020vls,Bea:2021zsu,Ecker:2021cvz} (see e.g.~\cite{CasalderreySolana:2011us} for a review). Here we will apply it to a four-dimensional gauge theory  
with a first-order, thermal phase transition. Imagine placing the theory in an initial homogeneous state with an energy density in the unstable, spinodal region (see \fig{fig:EvsT}). If this state is perturbed, the system will evolve to a final state that will necessarily be inhomogeneous. Following this real-time evolution with conventional quantum-field theoretical methods for an interacting gauge theory  is challenging. For this reason, Refs.~\cite{Attems:2017ezz,Janik:2017ykj,Attems:2019yqn} used holography, i.e.~they followed the evolution by solving the time-dependent Einstein's equations on the gravity side. The evolution consists of four generic stages \cite{Attems:2019yqn} visible in \fig{reflecting1}: A first, linear stage in which the instability grows exponentially; a second, non-linear stage in which peaks and/or phase domains are formed; a third stage in which these structures  collide and merge; and a fourth stage in which the system finally relaxes to a static, phase-separated configuration. In any given evolution of this type, the velocities of the different domains are a complicated function of the  perturbation of the initial, homogeneous state. This makes it difficult to perform a systematic study of the physics of the collision as a function of the domain velocities. In this paper we will overcome this difficulty by directly preparing initial states consisting of domains moving towards each other at a wide range of velocities. 
 

On the gravity side the above dynamics translates into the physics of a time-dependent, inhomogeneous horizon. Understanding this physics provides one motivation for our work. For example, we will examine in detail how the area of the horizon changes as two domains merge. Another motivation comes from heavy ion collision (HIC) experiments such as  the beam energy scan at RHIC, the compressed baryonic matter experiment at FAIR and other experiments at NICA. These experiments will open an unprecedented window into the properties of the phase diagram of Quantum Chromodynamics (QCD) at large baryon chemical potential, which is expected to contain a line of first order phase transitions ending at a critical point \cite{Stephanov:1998dy,Stephanov:1999zu,Stephanov:2017ghc}. If this scenario is realised then the real-time dynamics  of the spinodal instability may play an important role. 

One technical aspect that will allow us to go significantly beyond \cite{Attems:2017ezz,Attems:2019yqn} is our choice of scalar potential on the gravity side. The choice in those references gave rise to a phase diagram in the gauge theory with a large hierarchy between $\mathcal{E}_{high}$ and $\mathcal{E}_{low}$. This large ratio resulted in a run-time for our code on the scale of months. In contrast, here we will use the potential introduced in \cite{Bea:2018whf} and subsequently used in \cite{Bea:2020ees,Bea:2021zsu}. With this choice the ratio $\mathcal{E}_{high}/\mathcal{E}_{low}$ is of order one and the run-time decreases to minutes. 

We emphasize from the start that the domains we consider here should not be confused with the bubbles created via a nucleation process in first-order phase transitions. The expansion of those bubbles is driven by a pressure difference between the inside and the outside of the bubble. Instead, in our case the domains move simply because they are created with some initial velocity. In fact, if a number of bubbles are nucleated in a large -- but finite-size -- box, and if the average energy density in the box is between $\mathcal{E}_{high}$ and $\mathcal{E}_{low}$, then the endpoint of the evolution will be a phase-separated state with two domains with energies $\mathcal{E}_{high}$ and $\mathcal{E}_{low}$.


\section{The model}

Our gravity model is described by the Einstein-scalar action
\begin{equation}
\label{eq:action}
S=\frac{2}{\kappa_5^2} \int d^5 x \sqrt{-g} \left[ \frac{1}{4} {\cal R}  - \frac{1}{2} \left( \nabla \phi \right) ^2 - V(\phi) \right ]  \,.
\end{equation}
Exclusively for simplicity we assume that the scalar potential $V(\phi)$ can be derived from a superpotential $W(\phi)$ through the usual relation
\begin{equation}
 V(\phi)=-\frac{4}{3}W(\phi)^2+\frac{1}{2}W'(\phi)^2 \,.
\label{potentialsuperpotential}
\end{equation}
Different choices of (super)potential correspond to different dual four-dimensional gauge theories. As in \cite{Bea:2018whf,Bea:2020ees,Bea:2021zsu} we choose
\begin{equation}
\ell W(\phi)=-\frac{3}{2}-\frac{\phi^2}{2}-\frac{\phi^4}{4 \phi_M^2}+\frac{\phi^6}{\phi_Q}~,
\label{superpotential}
\end{equation}
where $\ell$ is the asymptotic curvature radius of the corresponding AdS geometry  and $\phi_M$, $\phi_Q$ are constants, which in this paper we will set to $\phi_M=1.0$ and $\phi_Q=10$. The dual gauge theory is  a conformal field theory (CFT) deformed with a dimension-three scalar operator with source $\Lambda$. This intrinsic scale in the gauge theory will set the characteristic scale of much of the physics of interest.  On the gravity side, $\Lambda$   appears as a boundary condition for the scalar $\phi$.  In the limit $\phi_Q \to \infty$ the sextic term is absent and the model reduces to that in \cite{Attems:2017ezz,Attems:2019yqn,Attems:2018gou}. The motivation for the choice here is simplicity: the dual gauge theory is a non-conformal theory with a first-order phase transition with a ratio 
$\mathcal{E}_{high}/\mathcal{E}_{low}$ of order one, as shown in \fig{fig:EvsT}, and the dual gravity solution is completely regular even at zero temperature. This model has $T_{critical}/\Lambda=0.3959$, $\mathcal{P}_{critical}/\Lambda^4=0.1124$, $\mathcal{E}_{high}/\Lambda^4=1.3736$ and $\mathcal{E}_{low}/\Lambda^4=0.4526$.

For simplicity, in this paper we will allow for dynamics only along one of the three spatial directions of the gauge theory, which we call $z$. In other words, we impose translational symmetry along $x,y$. We will also refer to $z$ as the longitudinal coordinate and to $x,y$ as the transverse directions. We will decompose the gauge theory stress tensor accordingly and work with the rescaled quantities 
\be
({\cal E}, \Pt, \Pt, \Pl, \mathcal{J})=\frac{\kappa_5^2}{2\ell^3} (-T^t_t, T^x_x, T^y_y, T^z_z, T^t_z) \,,
\ee
where ${\cal E}$ is the energy density, $\Pl$ and $\Pt$ are the longitudinal and transverse pressures, and $\mathcal{J}$ is the momentum density in the $z$-direction. For an $SU(\nc)$ gauge theory the prefactor on the right-hand side typically scales as $\nc^{-2}$.  In addition, we compactify the $z$-direction on a circle of length $L$. This infrared cut-off is technically convenient since it reduces the number of unstable modes of homogeneous states in the spinodal region to a finite number. In addition, compactifying the $z$-direction on a circle also brings about interesting new effects even for a single, boosted domain. We will start by  analyzing this case in Sec.~\ref{one} and then we will move to collisions of two domains in Sec.~\ref{collisions}. 
Note that we are slightly abusing the word ``domain'' in the following sense. At the end of the evolution in \fig{reflecting1} there are two domains with energy densities $\mathcal{E}_{high}$ and $\mathcal{E}_{low}$. Nevertheless, we will often use the word ``domain'' to refer specifically to the high-density phase. In this sense, we will sometimes think of the end state of \fig{reflecting1} as a domain surrounded by a low-density bath.

\section{Domains in motion}
\label{one}

In a domain collision, each of the participants is initially moving independently of the other, surrounded by a low-density bath. In this section we will study the dynamics of such a single domain. We will construct an initial, out-of-equilibrium state with a non-uniform fluid velocity in the $z$-direction and watch it evolve into a steady-state with uniform velocity. In addition to its intrinsic interest, this will also help us understand the construction of initial states in the case of domain collisions. 

\begin{figure}[h!!!]
\begin{center}
	\includegraphics[width=.65\textwidth]{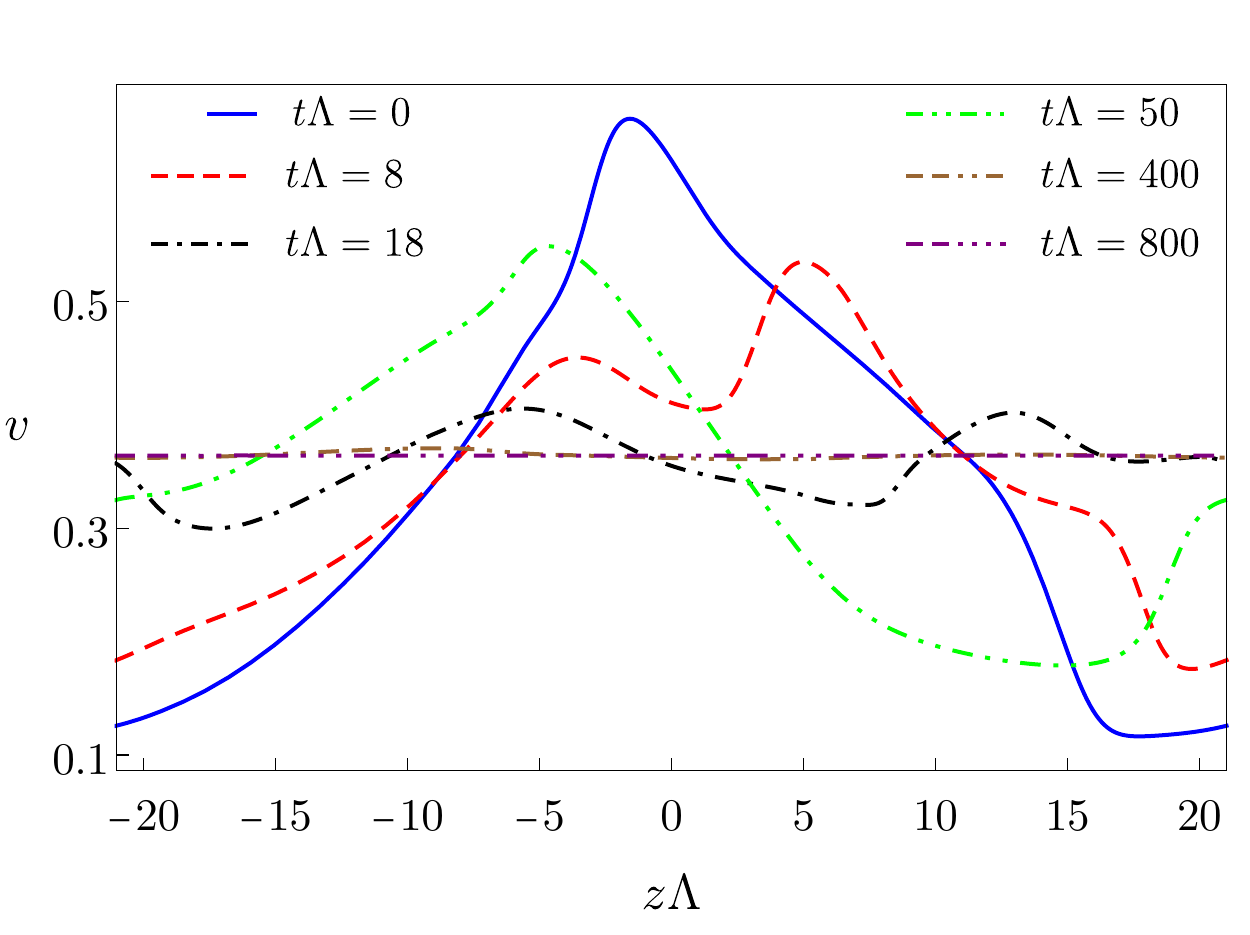} \\[2mm]
	\includegraphics[width=.70\textwidth]{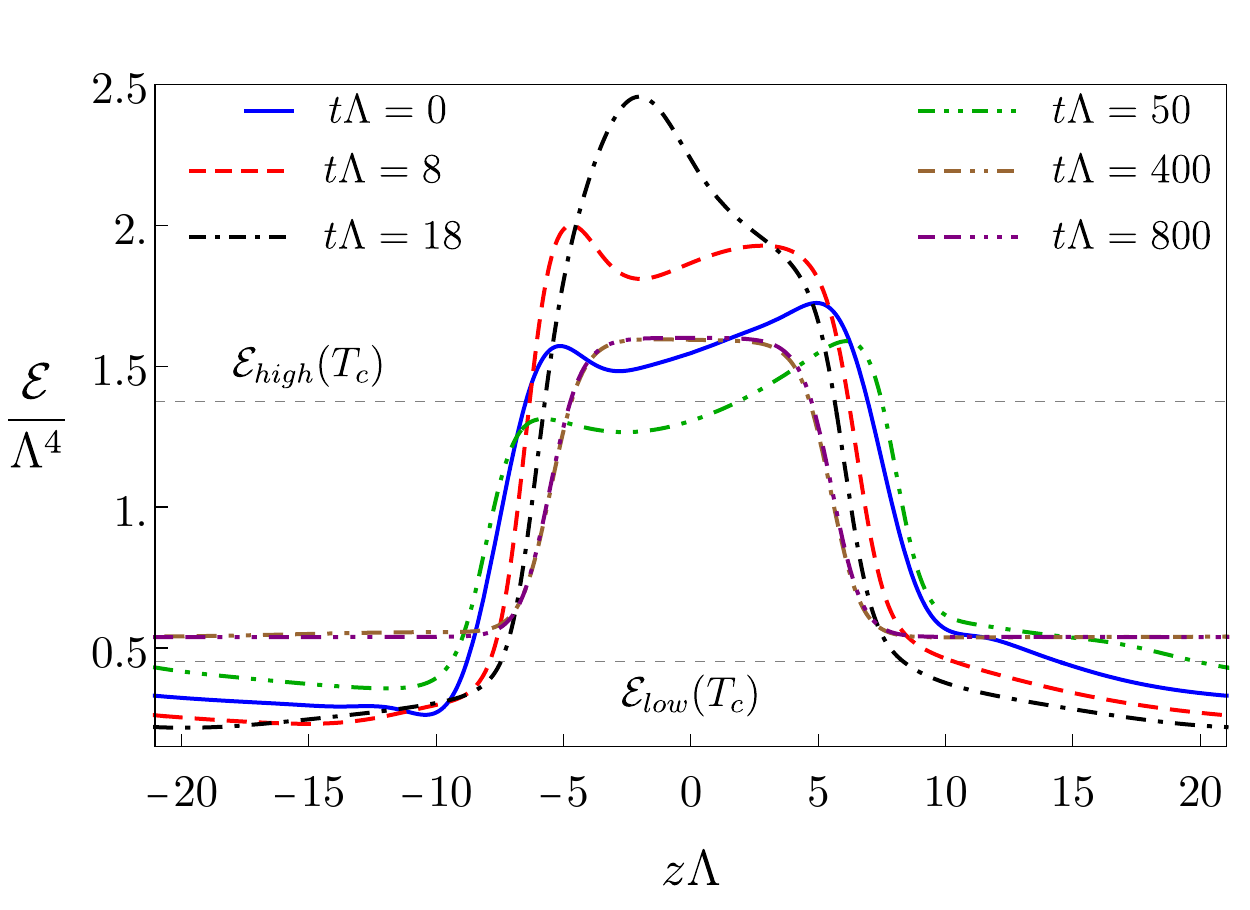} 
\end{center}
\vspace{-5mm}
\caption{\label{fig:transient} \small Time evolution of the fluid velocity (top) and of the energy density profile (bottom) for a domain in motion. To facilitate the comparison, each profile has been shifted by a constant amount to match the domain midpoint with $z\Lambda=0$. The maximum of the fluid velocity at $t\Lambda=0$ is 
$v_{max} = 0.66$.}
\vspace{4mm}
\end{figure}

We start with a single, static domain obtained as the end-state of a simulation like that in \fig{reflecting1}. We then modify the near-boundary fall-offs of the metric functions on the initial time slice on the gravity side\footnote{Specifically the subleading terms, which determine the expectation values.} so as to add a finite momentum density along the $z$-direction, $\mathcal{J}(z)$. We keep the total energy fixed as we inject this momentum; as we will see below this results in a decrease in the entropy of the system. The specific $z$-dependence is chosen to be 
\begin{equation}
\label{j}
\mathcal{J}(z)=j\, \Big(\mathcal{E}(z)-\mathcal{E}_{low}(T_c)\Big)\,,
\end{equation}
where $j$ is an appropriately chosen constant. At this initial step the momentum density has vanishing support on the low-density region. 
In order to be able to initialise our numerical code the initial value of $j$ cannot be too large. However, choosing a small value, letting the system evolve for a few time steps, increasing the value of $j$, and iterating this procedure, allows us  to eventually reach  high velocities. The reason for the iterations is that, after each change in the near-boundary fall-offs, a few time steps of evolution allow the bulk functions to adapt to the new near-boundary fall-offs. In particular, this allows our  code to find the new horizon within our numerical grid. We will define as the initial time, $t\Lambda=0$, the time at the end of this iterative process. At each step in the iteration process a fraction of the total momentum gets transferred to the low-density bath. As a consequence, at the end of the process the bath will not be completely at rest, as illustrated by the blue curve in  \fig{fig:transient}(top). Nevertheless, at this time most of the momentum density is still in the domain, where the fluid flow velocity is larger than in the bath. It is precisely this feature that tells us that the resulting state at $t\Lambda=0$ is out of equilibrium. 

\begin{figure}[h!!!]
\begin{center}
	\includegraphics[width=.72\textwidth]{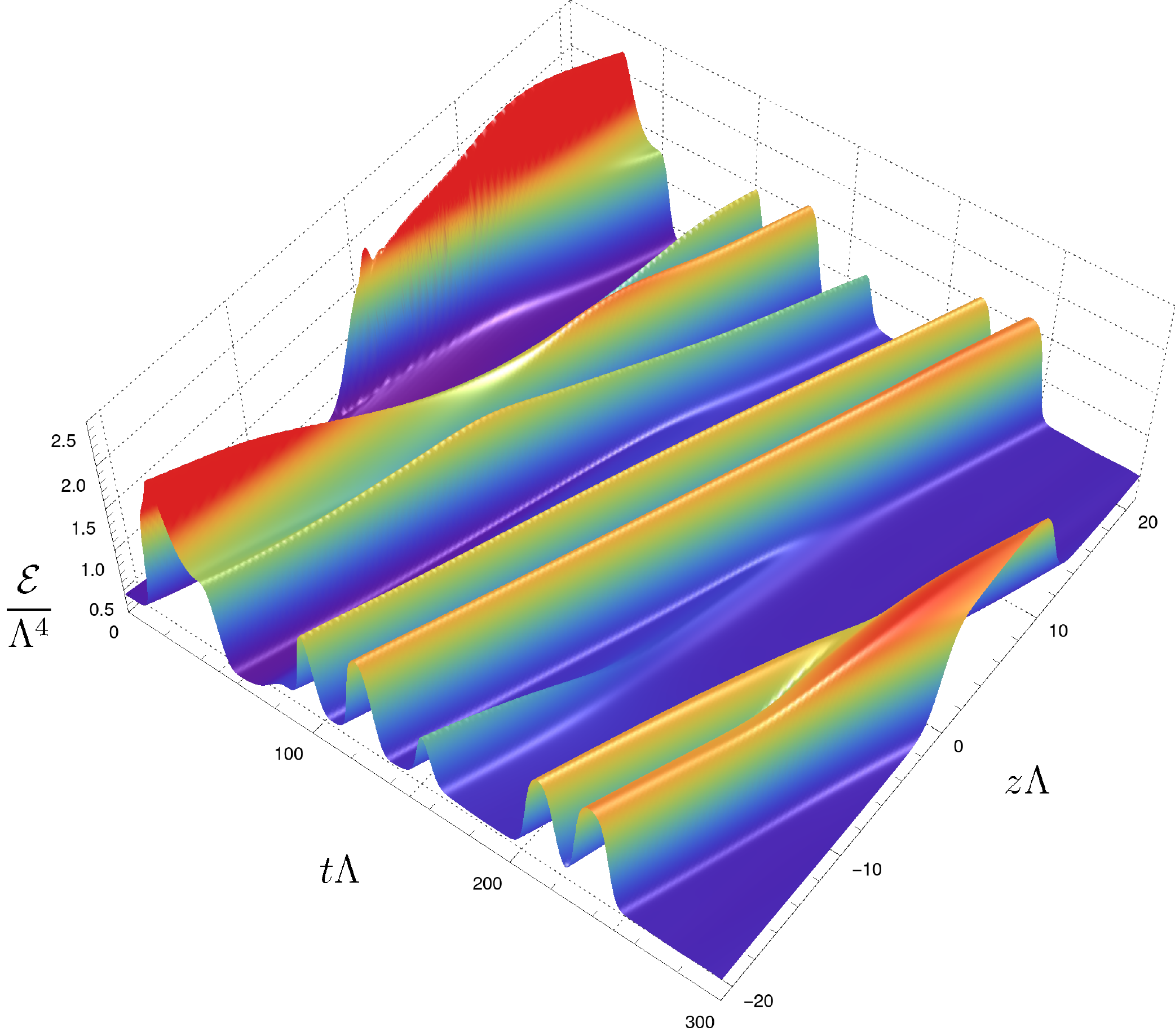} \\[2mm]
	\includegraphics[width=.72\textwidth]{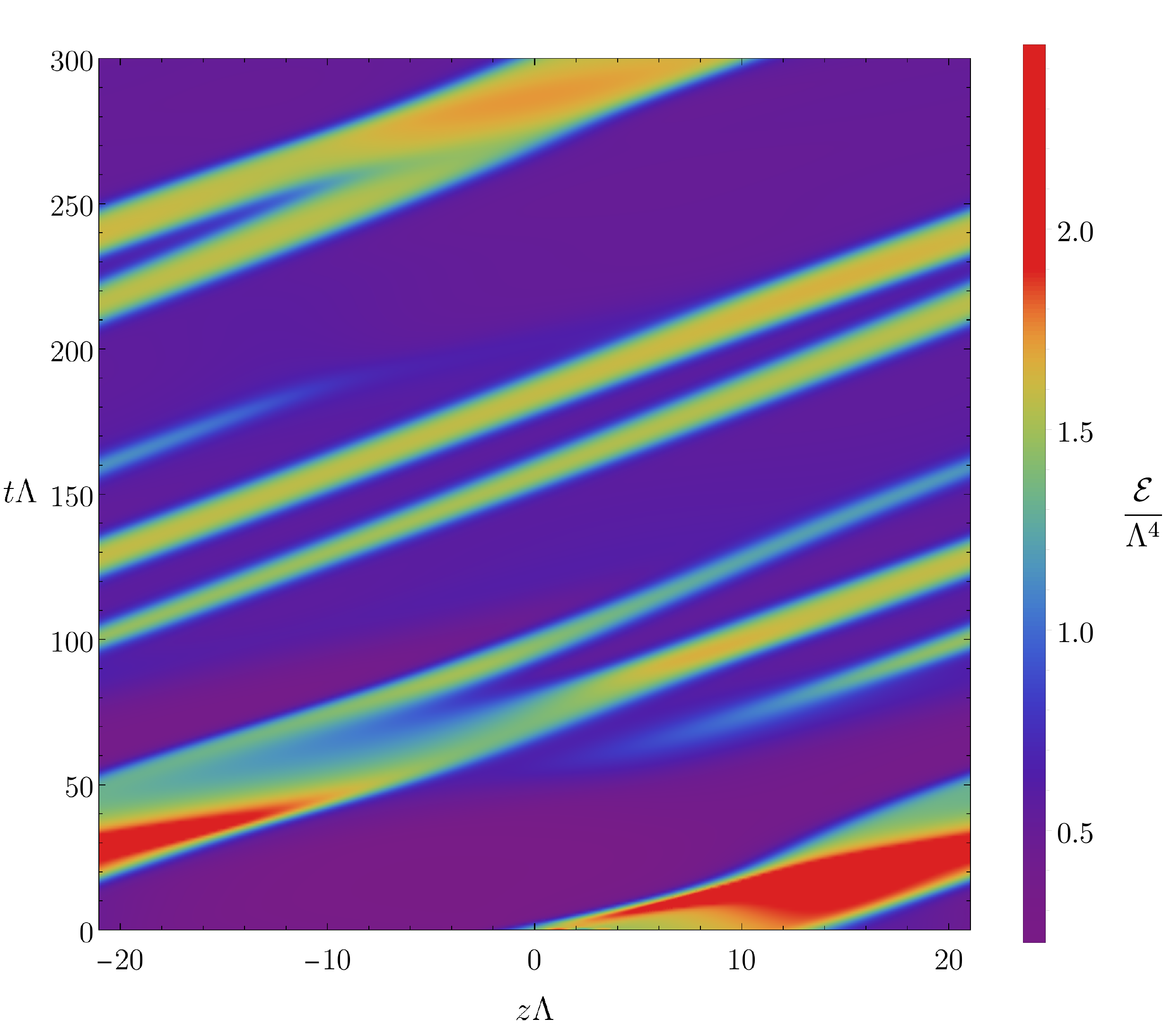}
\end{center}
\vspace{-5mm}
\caption{\label{sofar} \small Plots of the energy density for the evolution of an initial, single-domain state that splits into several fragments that subsequently merge back into a single domain.}
\vspace{7mm}
\end{figure}

\begin{figure}[h!!!]
\begin{center}
\begin{tabular}{cc}
	\includegraphics[width=.49\textwidth]{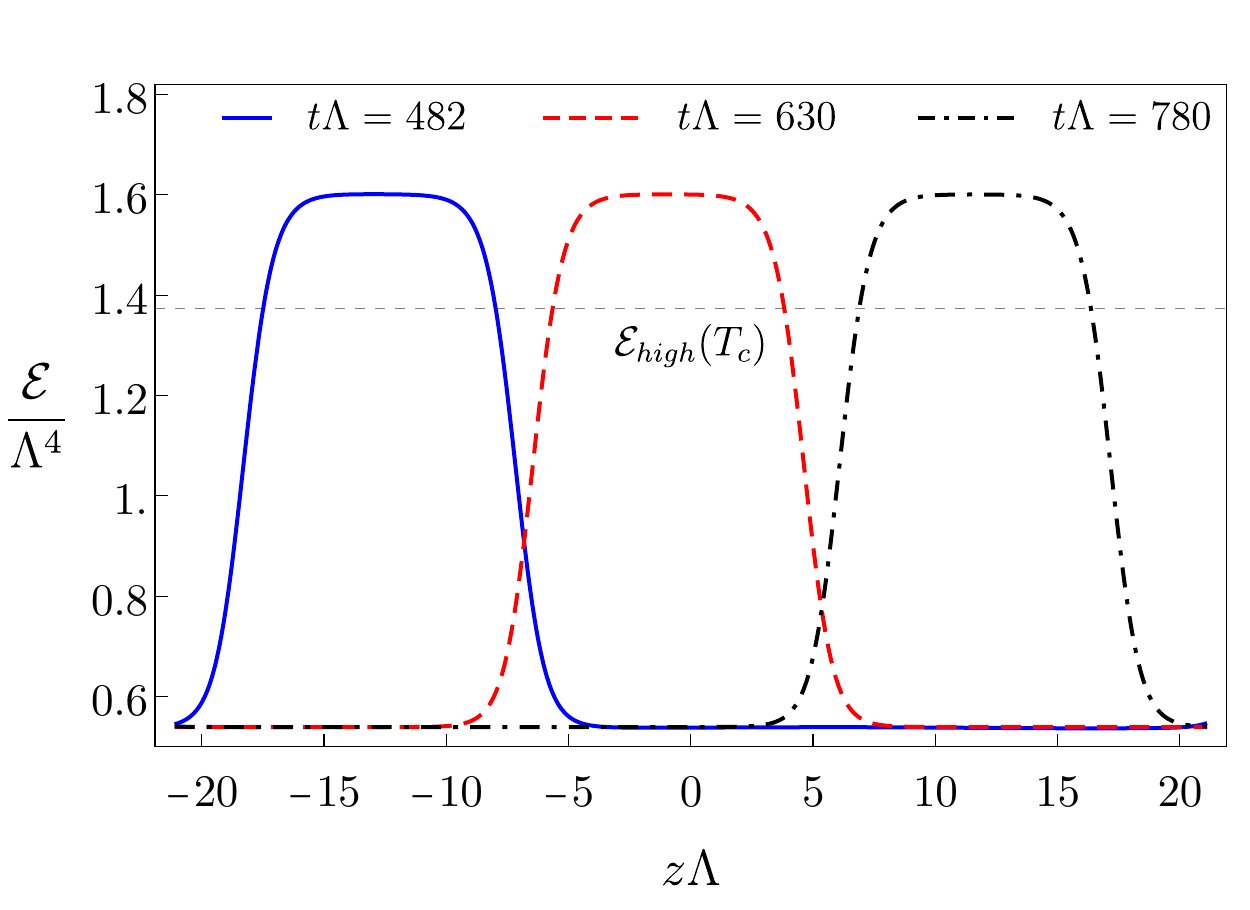} 
	\includegraphics[width=.49\textwidth]{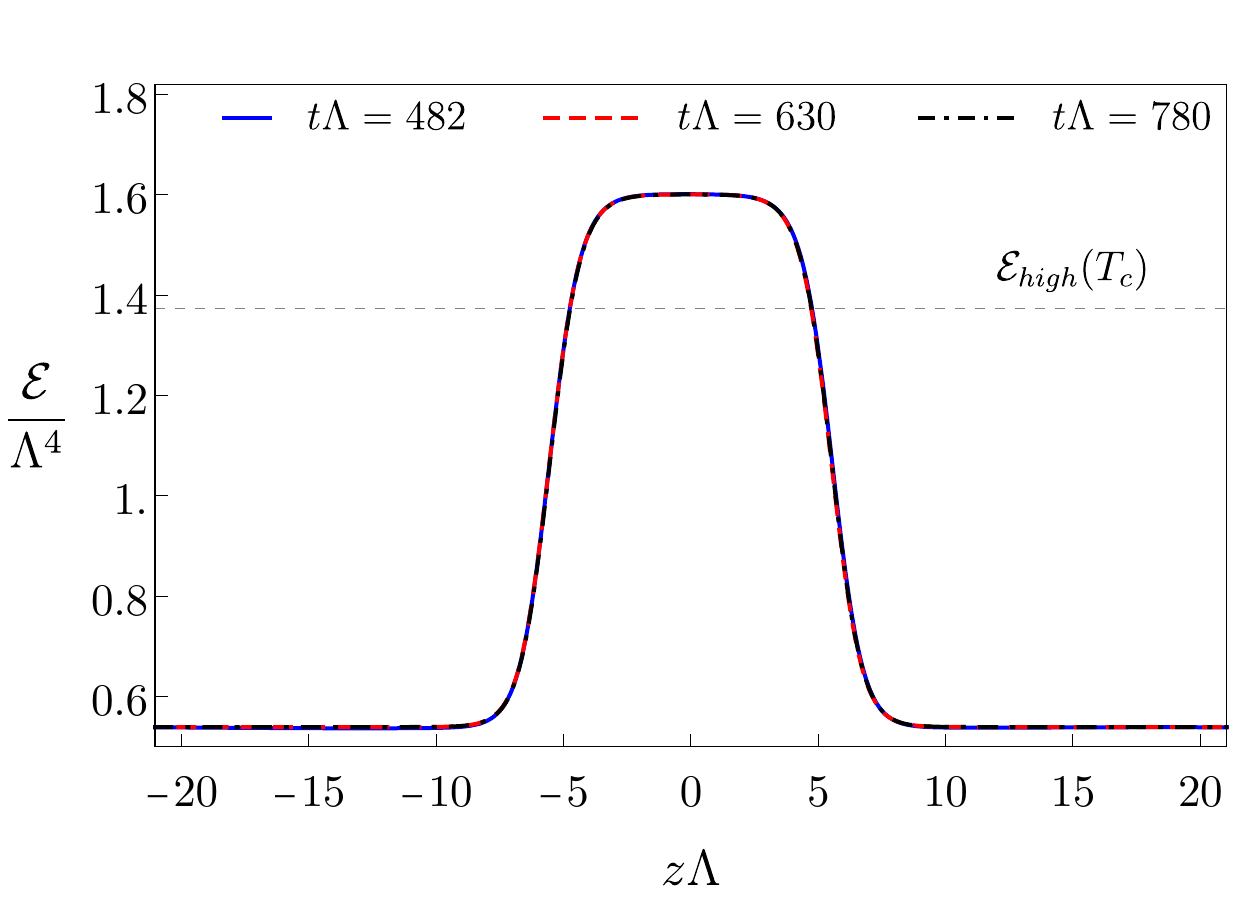} 
\end{tabular}
	\includegraphics[width=.55\textwidth]{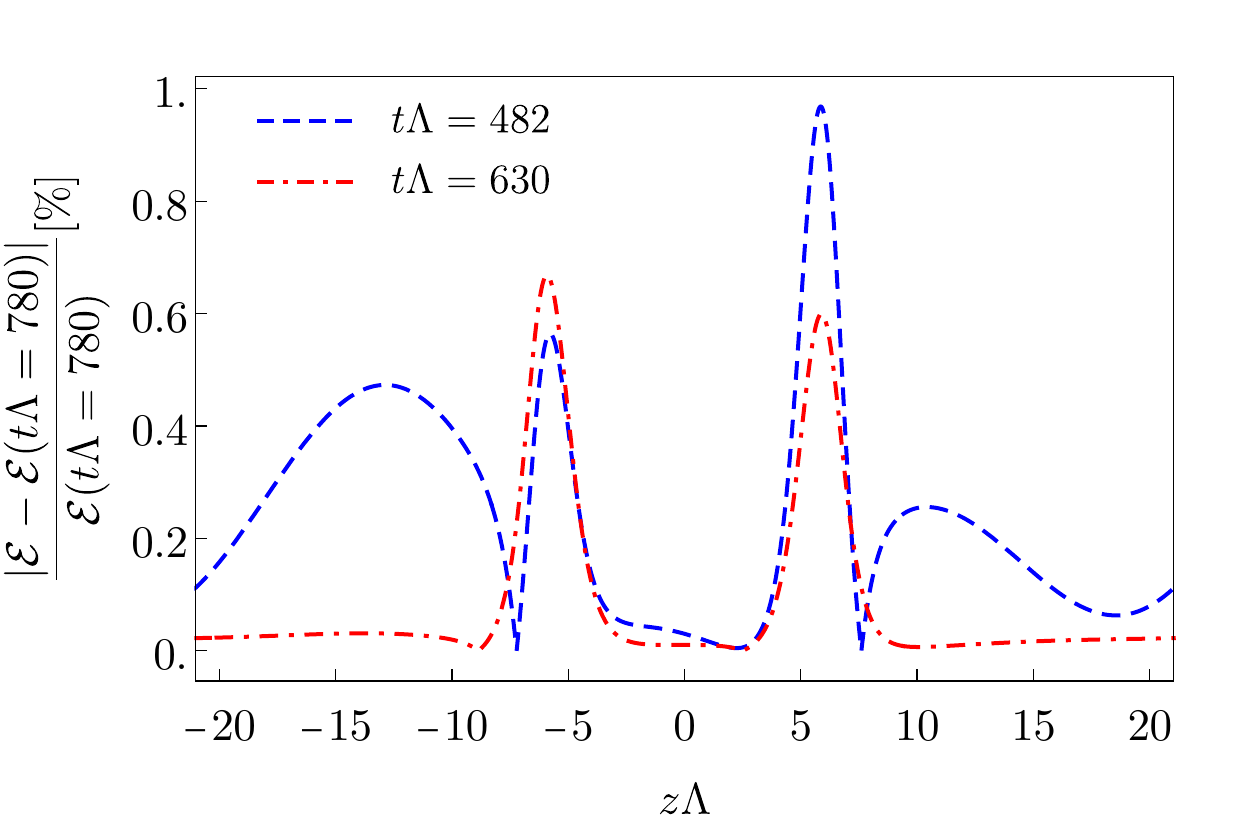} 
\end{center}
\vspace{-5mm}
\caption{\label{fig:stationary} \small (Left) Snapshots of the energy density at late times for the same initial state as in \fig{fig:transient}. (Right) The same profiles shifted by an appropriate constant. (Bottom) Relative difference between the profiles. }
\vspace{4mm}
\end{figure}

\begin{figure}[t]
\begin{center}
	\includegraphics[width=.6\textwidth]{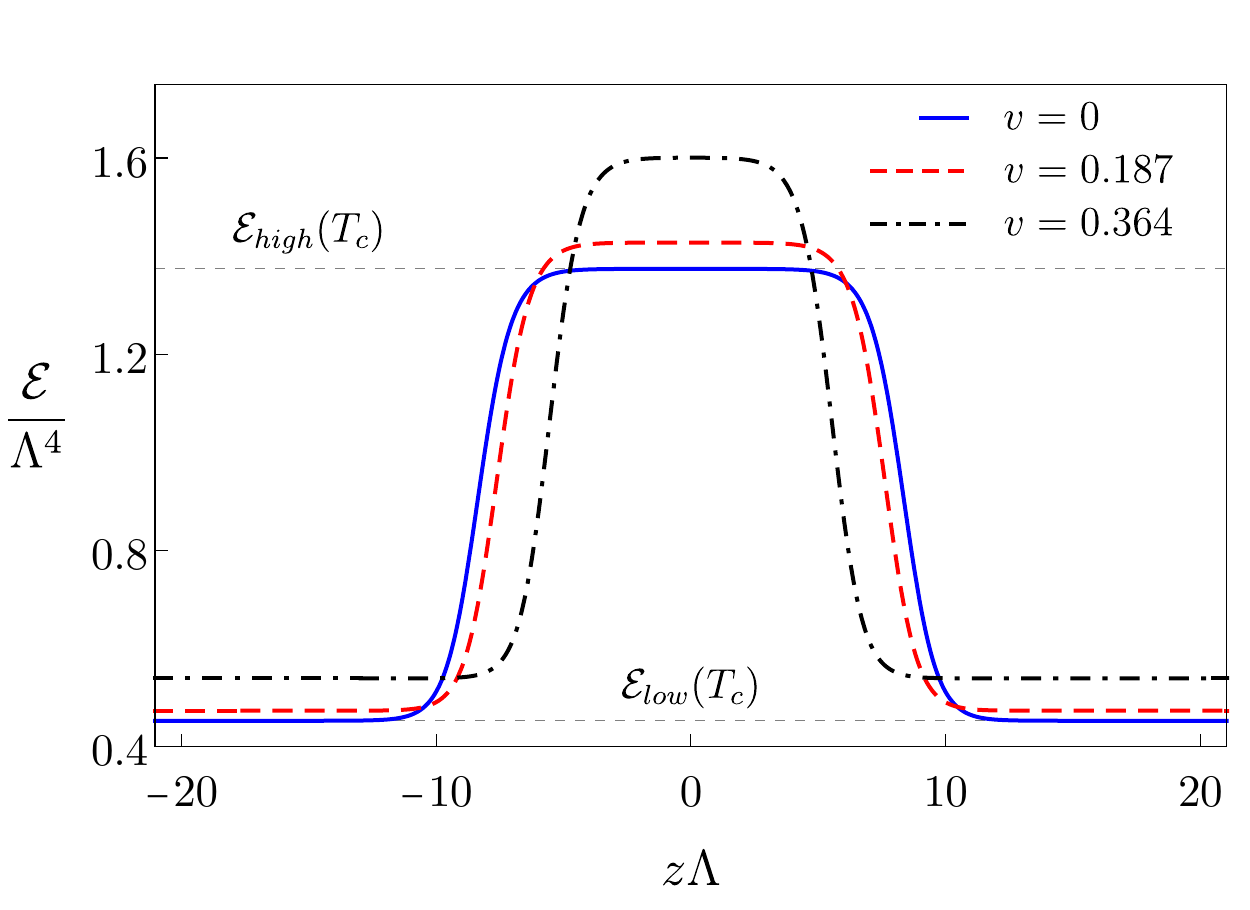} 
\end{center}
\vspace{-5mm}
\caption{\label{fig:domain_comparison} \small Steady-state energy profiles of domains with different velocities. The total energy in the box is the same in all cases. The average energy is  $\bar{\mathcal{E}}/\Lambda^4=0.817$ and the box has size $L\Lambda=42.0719$.}
\vspace{3mm}
\end{figure}

After $t\Lambda=0$ we do not inject any more momentum into the system. The subsequent evolution is characterized by a  significant energy and momentum transfer between the domain and the bath. The initial state is so far from equilibrium that it  splits into several fragments that eventually merge back to form a single domain, as illustrated in \fig{sofar}. 
The stationary state is reached around $t\Lambda\sim 400$. At this and at later times the fluid velocity field is constant along the $z$-direction with value  $v=0.364$, as seen in \fig{fig:transient}(top). In other words, the ``system moves as a whole''.  This is illustrated  in \fig{fig:stationary}, where we see that the energy density profile remains constant in time once the steady state is reached to a precision better than 1\%.  
This steady-state energy profile is narrower than that of the out-of-equilibrium state  at $t\Lambda=0$, as can be seen in \fig{fig:transient}(bottom), and it is also narrower than that of the static, equilibrium state with the same total energy, as illustrated in \fig{fig:domain_comparison} for two different velocities. 
\begin{figure}[h!!!]
\begin{center}
	\includegraphics[width=.6\textwidth]{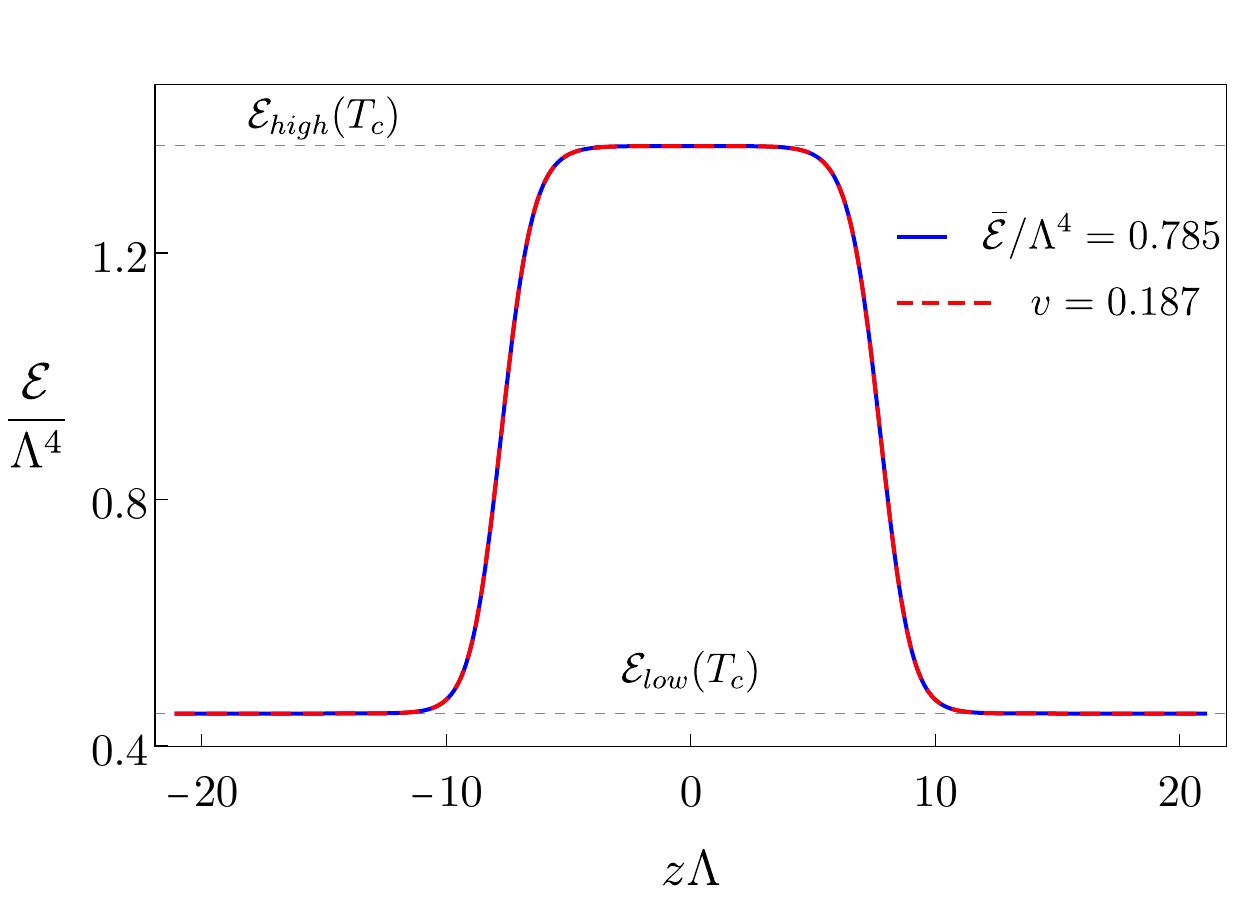}
\end{center}
\vspace{-5mm}
\caption{\label{fig:commoving-LAB} \small Comparison of the energy profile obtained by Lorentz-transforming the curve in \fig{fig:domain_comparison} (dashed, red curve) with a configuration obtained directly as an inhomogeneous, static configuration with the same average energy (\ref{average}) in a box of size $L'=\gamma L$ (solid, blue curve).}
\vspace{7mm}
\end{figure}

We see in the figure that the domains become narrower as the velocity increases, and that the energy densities in the domain and in the bath are higher than 
$\mathcal{E}_{high}(T_c)$ and $\mathcal{E}_{low}(T_c)$, respectively.  These features are consistent with the fact that, by construction, all these configurations have the same total energy. 

In the steady state, a comoving observer traveling with velocity $v$ sees a fluid configuration at rest. This observer uses coordinates $t', z'$ related to $t,z$ through  the usual Lorentz transformation 
\be
\label{lo}
t' = \gamma (t-vz) \sac z' = \gamma(z- vt) \sac \gamma = (1-v^2)^{-1/2}\,.
\ee
The original identification $(t,z) \sim (t, z+L)$, translates into a non-trivial identification for the comoving observer
\be
\label{ide}
(t',z') \sim (t'-\gamma v L, z' + \gamma L) \,.
\ee
In other words, for the co-moving observer the identification is not purely along the spatial direction but it involves time as well. This imposes a non-trivial constraint on any physical configuration seen by this observer, since any such configuration must be invariant under (\ref{ide}). However, for  configurations that are static with respect to the comoving observer this condition reduces to periodicity along the $z'$-direction with period $L'=\gamma L$. This is the case for the fluid configuration obtained by applying the Lorentz transformation (\ref{lo}) to the configurations of \fig{fig:domain_comparison}. In particular, the energy density in the comoving frame takes the form 
\be
\label{eprime}
\mathcal{E}' = \gamma^2\left( \mathcal{E} + v^2 P_z -2v \mathcal{J} \right) \,,
\ee
and the average energy in this frame is 
\be
\label{average}
\bar{\mathcal{E}}'=\bar{\mathcal{E}}-v \bar{\mathcal{J}} \,.
\ee
The result of applying \eqn{eprime} to the $v=0.187$ case of \fig{fig:domain_comparison} is shown by the dashed, red curve in \fig{fig:commoving-LAB}.
As expected, this matches the solid blue curve, corresponding to a profile obtained directly as an inhomogeneous, static configuration with average energy (\ref{average}) in a box of size $L'$.

We see that, in the case of a single domain, we could have obtained the steady states of \fig{fig:domain_comparison} by applying the inverse Lorentz transformation to the corresponding static configurations. In contrast, in the case of two domains moving towards each other that we will consider in the next section, it will be crucial to be able to construct the moving domains via the method described around \eqn{j}, because in that case we will need to match two baths with opposite velocities.

\section{Domain collisions}
\label{collisions}
\begin{figure}[t]
\begin{center}
\begin{tabular}{cc}
	\includegraphics[width=.49\textwidth]{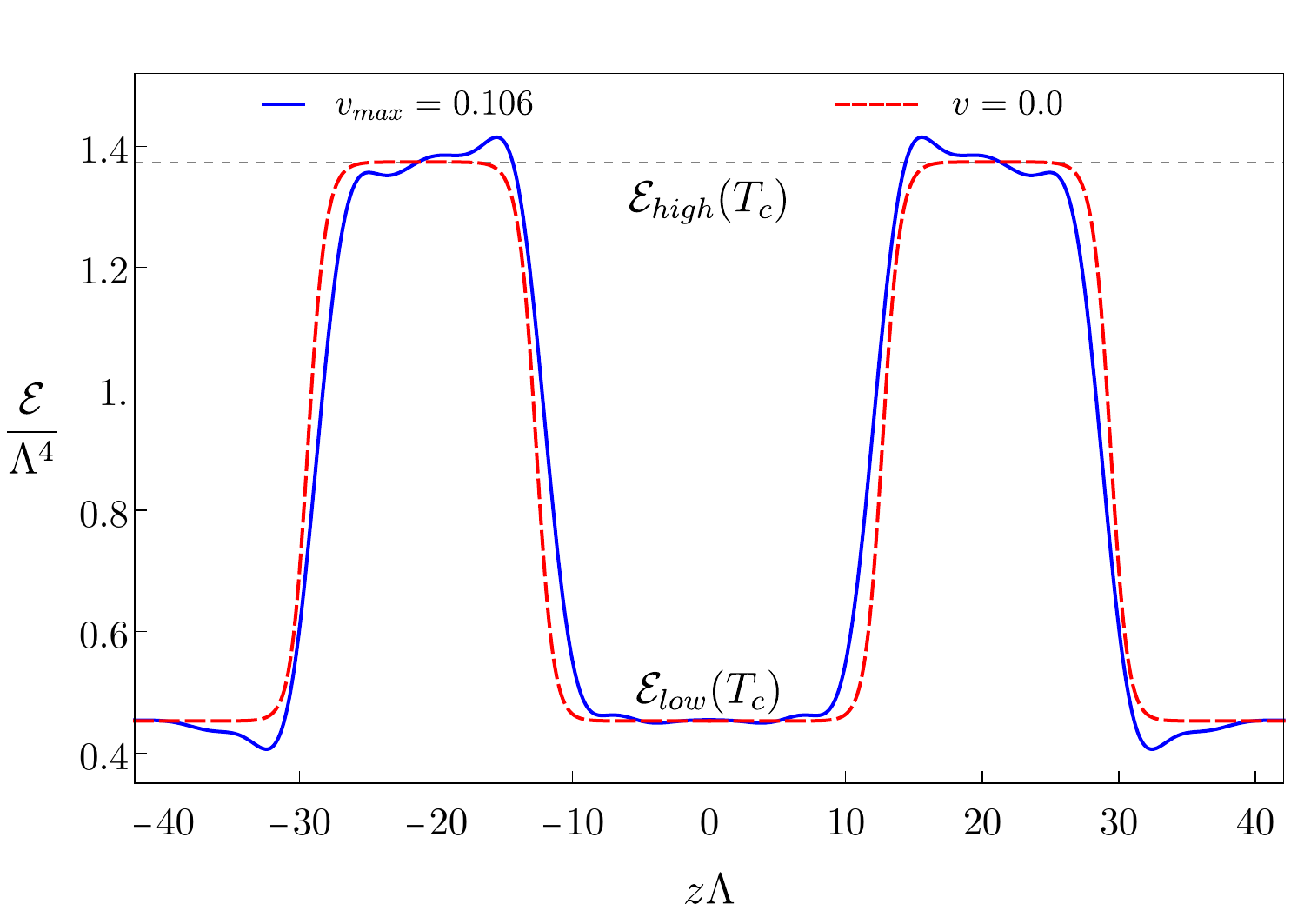} 
	\includegraphics[width=.49\textwidth]{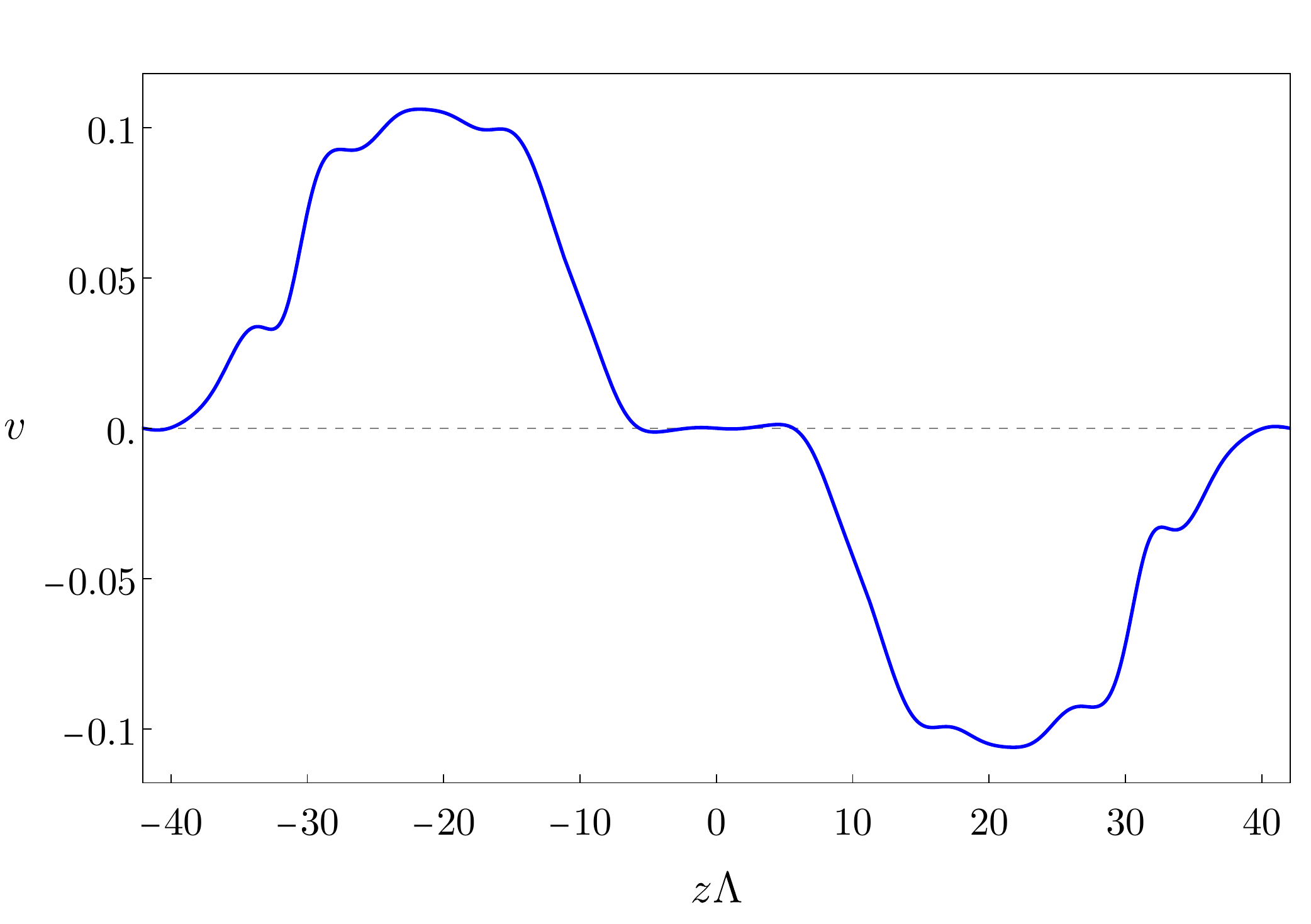}
\end{tabular}
\end{center}
\vspace{-5mm}
\caption{\label{fig:merging_initial} \small (Left) Initial configuration (dashed red) used to generate two domains in motion (solid blue) that will collide with each other.  (Right) Fluid velocity for the domains in motion on the left.}
\vspace{7mm}
\end{figure}

We now turn to the construction of initial states whose evolution will result in the collision of two domains. We start with a configuration consisting of two static domains separated by some distance from one another, like the one described by the dashed, red curve in \fig{fig:merging_initial}(left). This can be constructed, for example, by gluing together two individual domains along the low-energy bath. We place the two domains at antipodal points of the $z$-circle, in which case  the resulting configuration is unstable but static \cite{Attems:2019yqn}. If this configuration is slightly perturbed then, upon time evolution, the domains will approach each other and eventually collide. However,  this collision will typically happen at low relative velocity \cite{Attems:2019yqn}. Therefore, in order to explore the collision physics over a significant range of relative velocities, we will modify the initial configuration following the procedure described around \eqn{j}. The crucial difference is that now we inject  momentum densities of equal magnitudes but opposite signs in each domain. As in the case of a single domain, at the end of the iteration procedure some of the momentum is contained in the bath but most of it is in the domains, as illustrated in \fig{fig:merging_initial}(right). Also, the shape of the domains, described by the solid,  blue curve in \fig{fig:merging_initial}(left), is modified with respect to the static situation. Note that, by symmetry, the fluid velocity must vanish at the two middle points in between the domains, i.e.~at $z\Lambda=0$ and $z\Lambda=L/2$. This is the reason why this initial state cannot be obtained by gluing together two individual steady states of the previous section moving towards each other, since at the middle points the baths would be moving with non-zero, opposite velocities, and hence the velocity field would be discontinuous. One may think that an alternative method would be to use superposition instead of gluing. Specifically, we recall that, in the characteristic formulation that we use on the gravity side (see \cite{Attems:2017zam} for details), the  data on the initial time slice is free. We can therefore superpose the initial data corresponding to two individual domains moving towards each other. Although this procedure is conceptually sensible, it is technically challenging to find the apparent horizon in the resulting geometry. For this reason we followed the procedure described around \eqn{j}. 

We performed a series of identical simulations except for the fact that  we varied the maximum velocity of the fluid in the initial state  in the range $v_{\mathrm{max}}\in (0,0.73)$. We use this velocity as one possible characterization of the initial state. In all cases we fixed the distance between the mid points of the domains to $d\Lambda\simeq 42$. 
We found three qualitatively different dynamical regimes. For low values $v_{\mathrm{max}} \lesssim v_1 \simeq 0.09$  the domains initially slow down, enter a quasi-static regime, and eventually collide and merge. For intermediate values $v_1 \lesssim v_{\mathrm{max}} \lesssim v_2\simeq 0.2$ the quasi-static phase is absent.  For high values $v_{\mathrm{max}} \gtrsim v_2$ the domains collide but do not merge. Instead they break apart and merge only in subsequent collisions. 

We emphasize that there is no sharp transition between these regimes and hence the differences are qualitative in nature. For the specific simulations with the domains of \fig{fig:merging_initial} we found the quoted values $v_1 \simeq 0.09, v_2 \simeq 0.2$. Repeating these simulations for different initial domains we found that the values of $v_1, v_2$ depend on the choice of initial domains. However, the existence of three qualitatively different regimes persisted, suggesting that it is a robust property of the collision dynamics. 



\subsection{Low velocity}

\begin{figure}[h!]
\begin{center}
	\includegraphics[width=.65\textwidth]{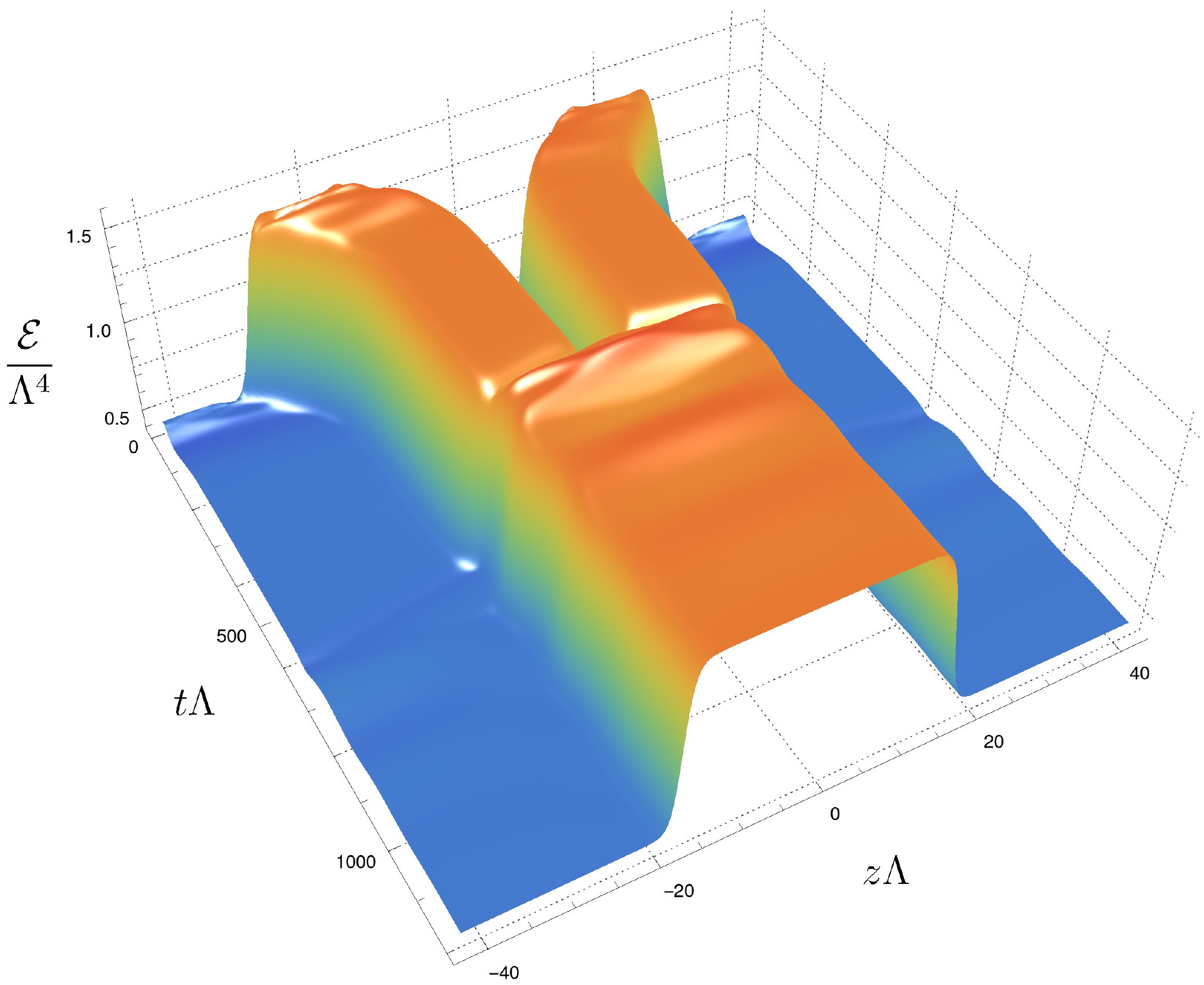} \\[2mm]
	\includegraphics[width=.65\textwidth]{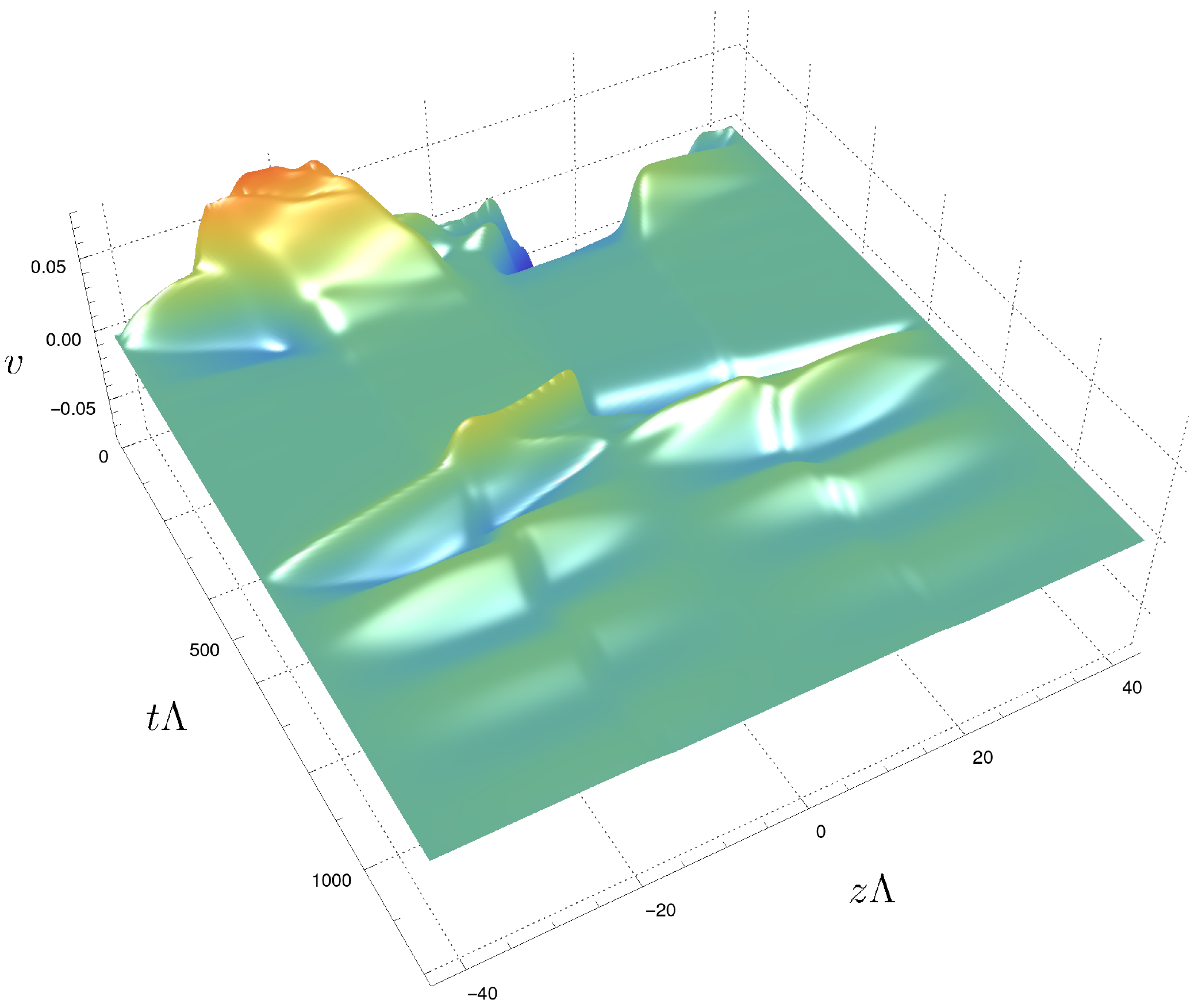} 
\end{center}
\vspace{-5mm}
\caption{\small
Evolution of the energy density (top) and of the fluid velocity (bottom) for a simulation with $v_{\mathrm{max}}= 0.08$. }
\vspace{1mm}
\label{fig:3D_vmax005}
\end{figure}

\fig{fig:3D_vmax005} shows the evolution of the energy density and the fluid velocity for a simulation with $v_{max}=0.08$. We can distinguish several stages of the evolution: the initial slowing down of the domains, the quasi-static regime, the merging into a single domain and the subsequent relaxation to equilibrium.

\begin{figure}[h!!!]
	\begin{center}
		\begin{tabular}{cc}
			\includegraphics[width=.49\textwidth]{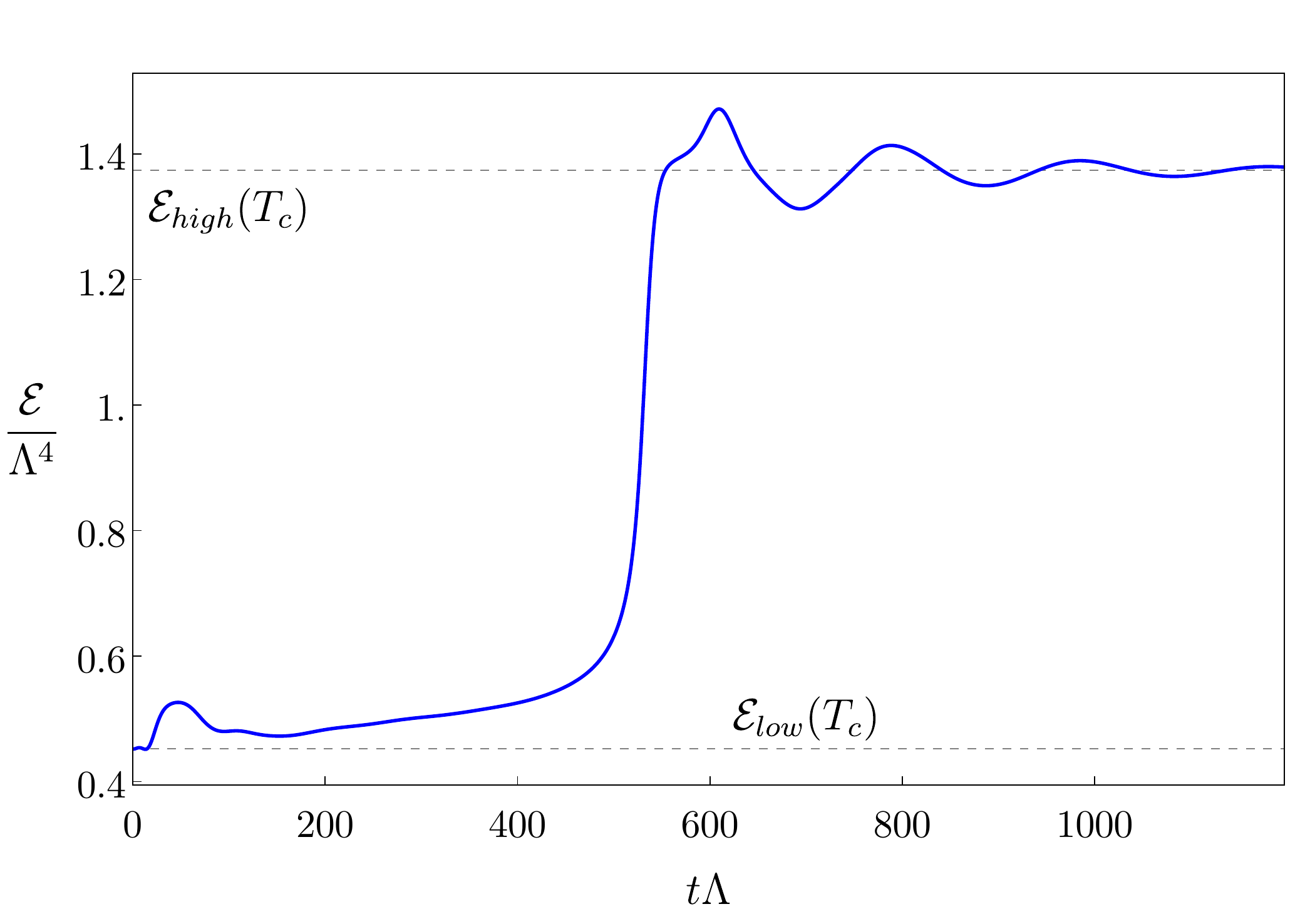} 
			\includegraphics[width=.49\textwidth]{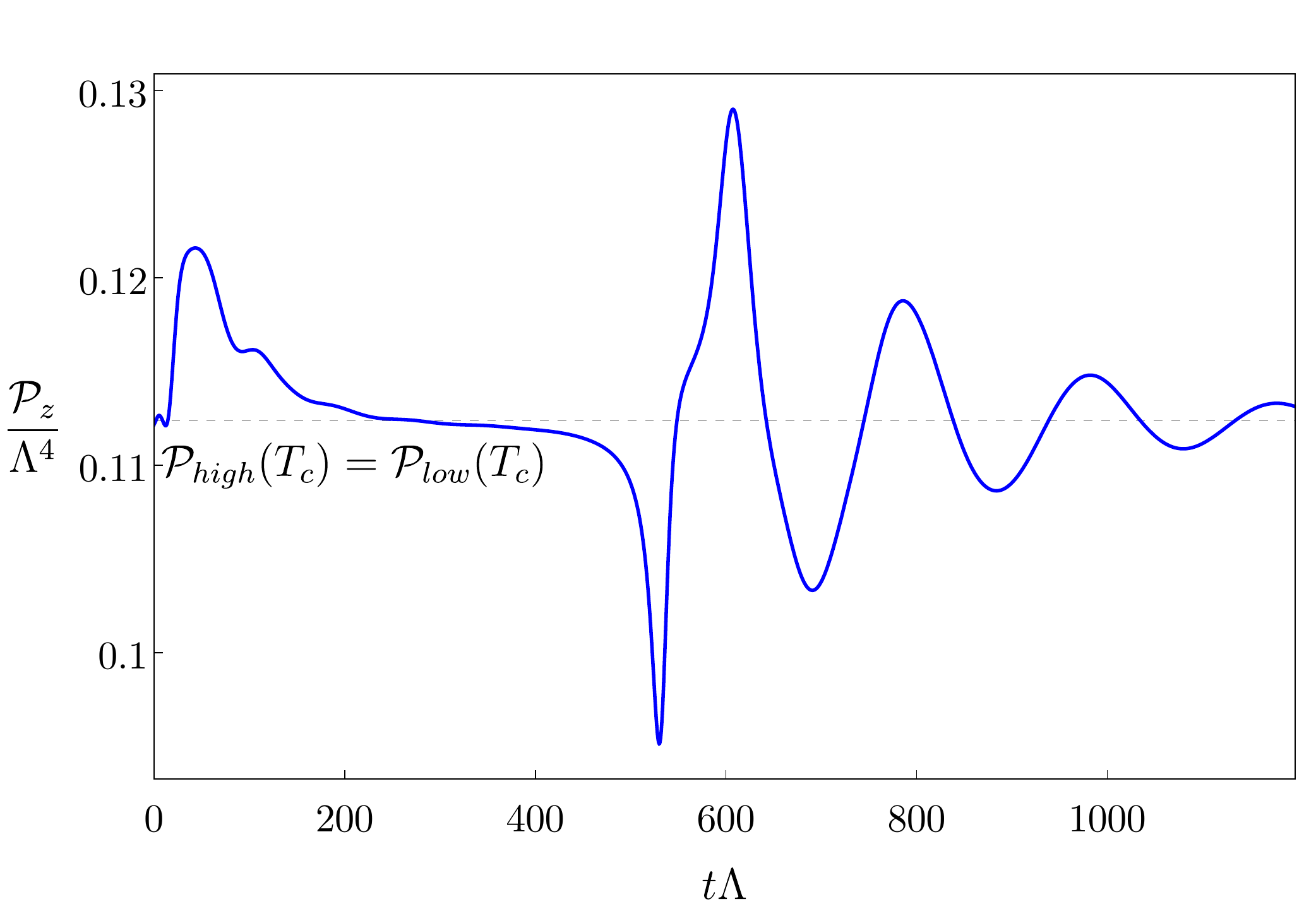} 
		\end{tabular}
	\end{center}
	\vspace{-5mm}
	\caption{\label{fig:E_P_middle_low_005} \small 
	Energy density (left)  and longitudinal pressure (right) as a function of time at $z\Lambda=0$ for the collision of \fig{fig:3D_vmax005}.}
	\vspace{2mm}
\end{figure}

\begin{figure}[h!!!]
\begin{center}
	\includegraphics[width=.55\textwidth]{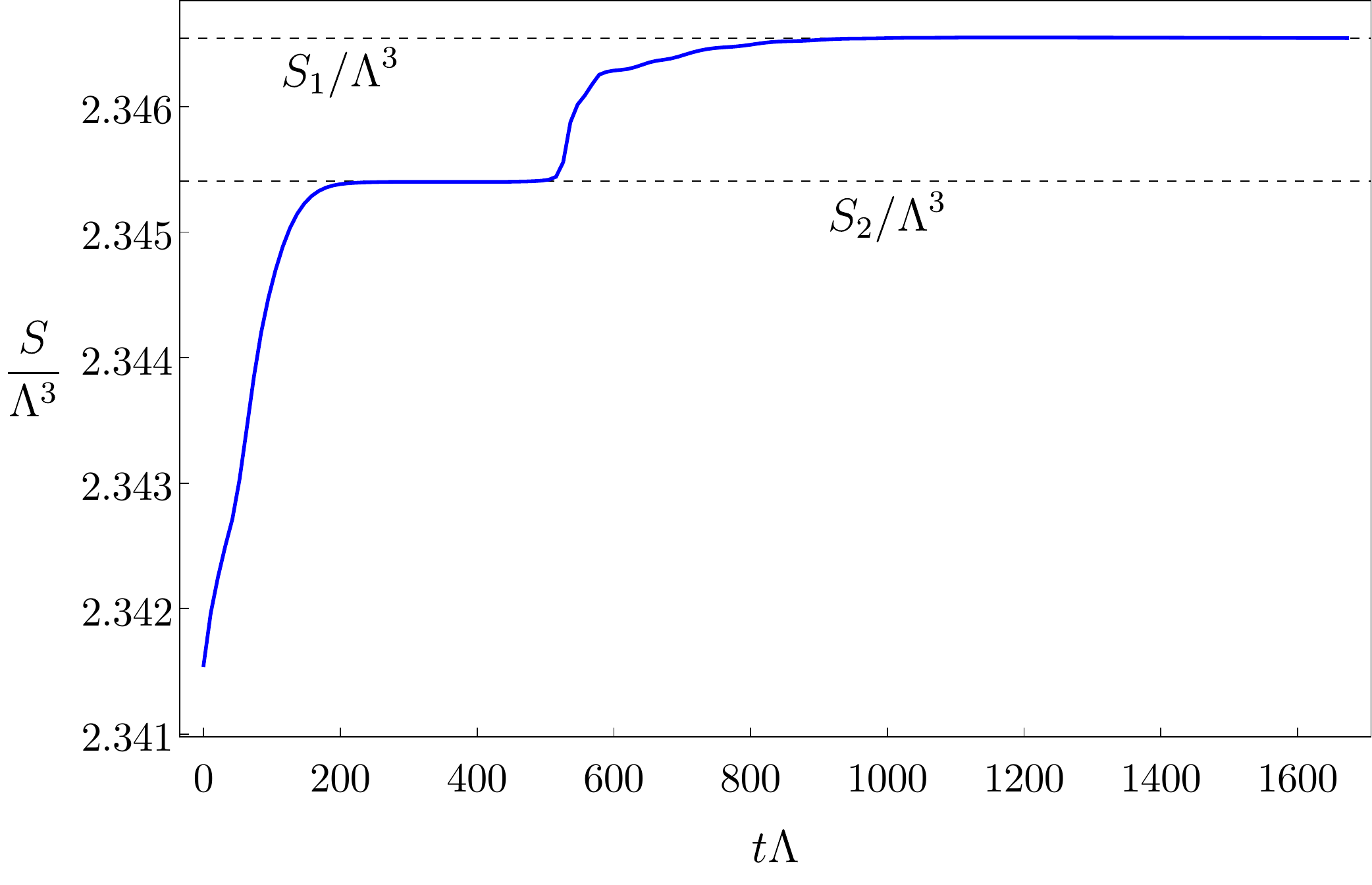} 
\end{center}
\vspace{-5mm}
\caption{\label{fig:vmax_005_S} \small Time evolution of the total entropy (per unit transverse area) for the collision of \fig{fig:3D_vmax005}. $S_2$ is the entropy of a static configuration with two domains (the dashed, red curve in \fig{fig:merging_initial}). $S_1$ is the entropy of a static configuration with the same total energy but a single domain.}
	\vspace{2mm}
\end{figure}

In the first stage, $t\Lambda\lesssim 100$, the domains slow down as they transfer momentum to the low-energy bath  in between them. This results in an increase of the energy density and of the  longitudinal pressure at the center, as shown at early times in \fig{fig:E_P_middle_low_005}. We use the area density of the apparent horizon on the gravity side as a proxy for the entropy density of the system. As shown in \fig{fig:vmax_005_S}, the dissipation in the initial phase causes  a considerable increase in the entropy of the system, meaning that it is comparable to the entropy increase in the final merger.  

\begin{figure}[h!!!]
	\begin{center}
		\begin{tabular}{cc}
			\includegraphics[width=.49\textwidth]{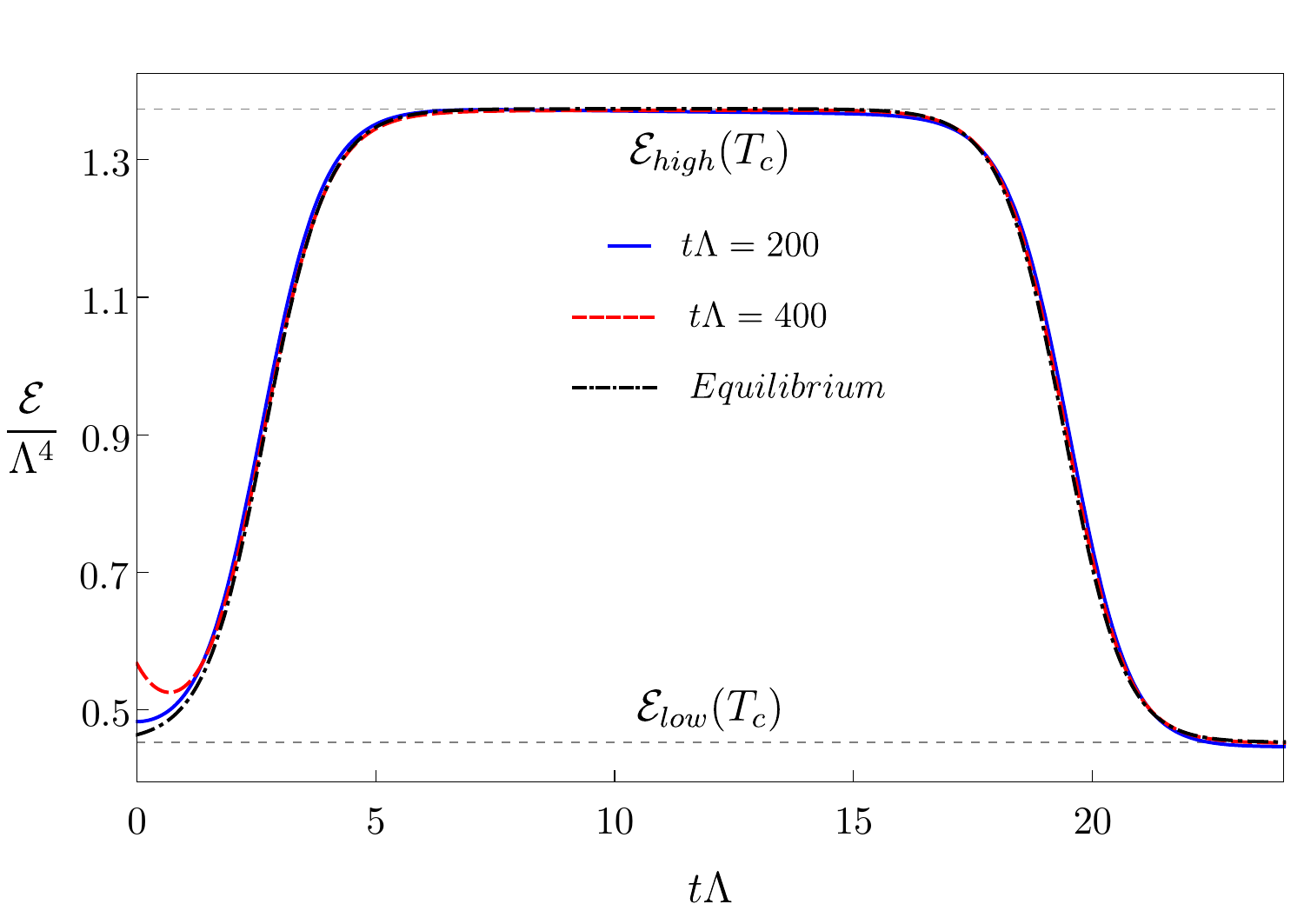} 
			\includegraphics[width=.525\textwidth]{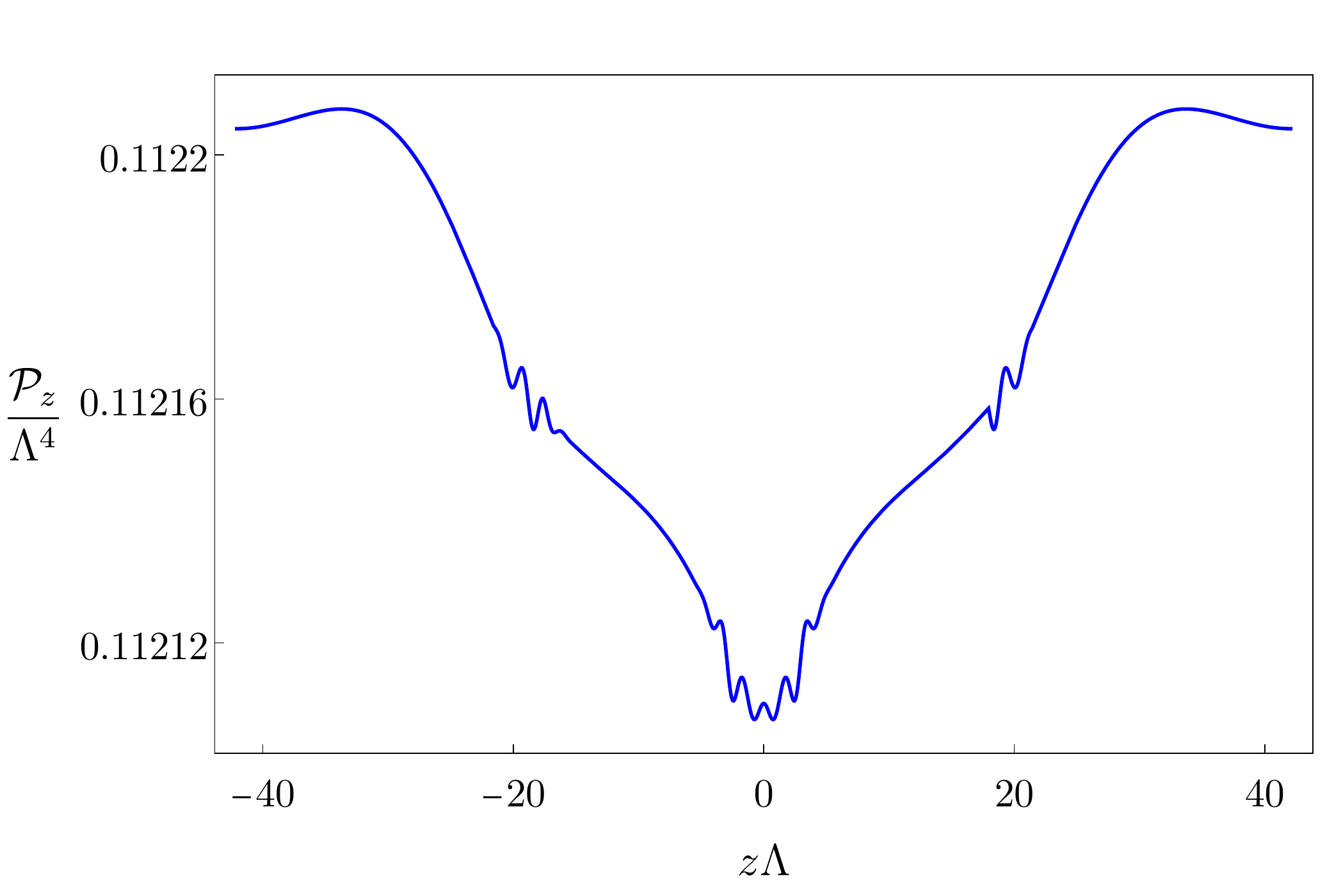} 
		\end{tabular}
	\end{center}
	\vspace{-5mm}
	\caption{\label{fig:E_domains_comparison} \small (Left) Energy profiles (shifted by an appropriate amount) of the domains during the quasi-static phase for the collision of \fig{fig:3D_vmax005}. (Right) Snapshot of the longitudinal pressure distribution at $t\Lambda=350$.}
	\vspace{2mm}
\end{figure}

In the period $100 \lesssim t\Lambda \lesssim 500$ the energy density (the pressure) at the centre increases (decreases) slowly. The quasi-static nature of this regime can be clearly seen in the entropy, which displays a plateau that extends up to $t\Lambda \sim  500$. Note that the entropy of the plateau coincides with that of the initial two-domain configuration (dashed, red curve in \fig{fig:merging_initial}(left)), which is static. Presumably, this agreement is due to the fact that the velocity of the two domains in the quasi-static regime is extremely low, as can be seen in \fig{fig:3D_vmax005}(bottom). This is consistent with the fact that the domains move as rigid bodies whose profiles are very well approximated by the equilibrium profile, as shown in \fig{fig:E_domains_comparison}(left). The right plot in this figure also shows that the dynamics is driven by a small pressure deficit in the region in between the domains. 

The fact that the entropy of the system at times $t\Lambda \lesssim 100$ is lower than the entropy of two static domains means that 
the procedure by which we inject momentum into the system at constant total energy decreases the entropy. This can be easily understood on the gravity side, where this procedure increases the total momentum of the dual black brane while keeping its mass fixed. If one performs a Kaluza-Klain reduction along the $z$-direction, this corresponds to increasing the electric charge of the resulting, lower-dimensional  black brane while keeping its mass fixed. Just in a  Reissner-Nordstr\"om black hole, this procedure is expected to reduce the entropy.


\begin{figure}[t]
\begin{center}
	\includegraphics[width=.8\textwidth]{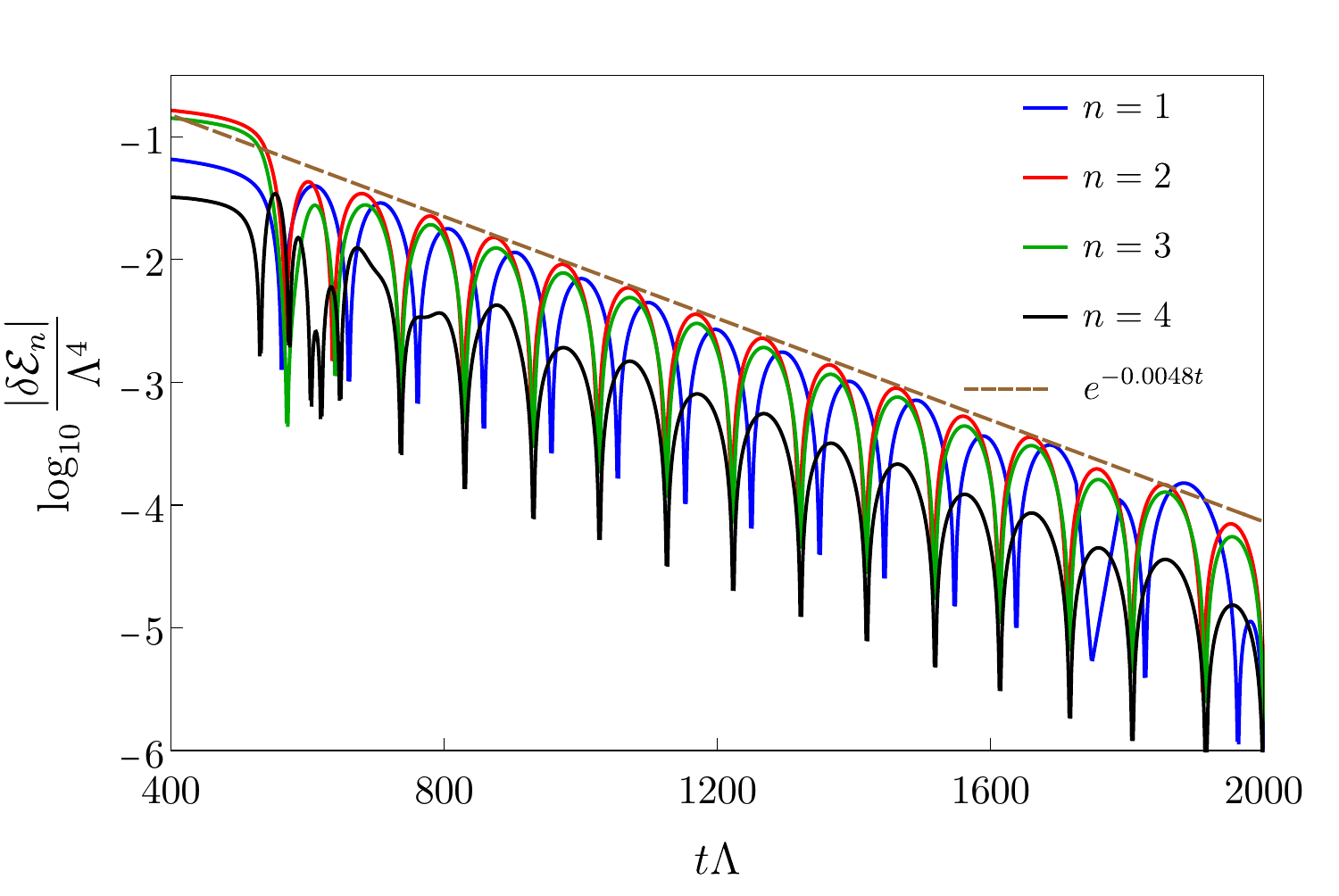} 
\end{center}
\vspace{-5mm}
\caption{\label{fig:Fourier_005}  Time evolution of the spatial Fourier modes of $\delta\mathcal{E}=\mathcal{E}(t,z)-\mathcal{E}_{eq}(z)$ after the merging. All modes evolve as $\delta\mathcal{E} _n\sim\exp(\omega_1 t)\cos(\omega_2 t)$ with the same $(\omega_1,\omega_2)=(-0.012,0.063)$.}
	\vspace{7mm}
\end{figure}

\begin{figure}[t]
	\begin{center}
		\begin{tabular}{cc}
			\includegraphics[width=.49\textwidth]{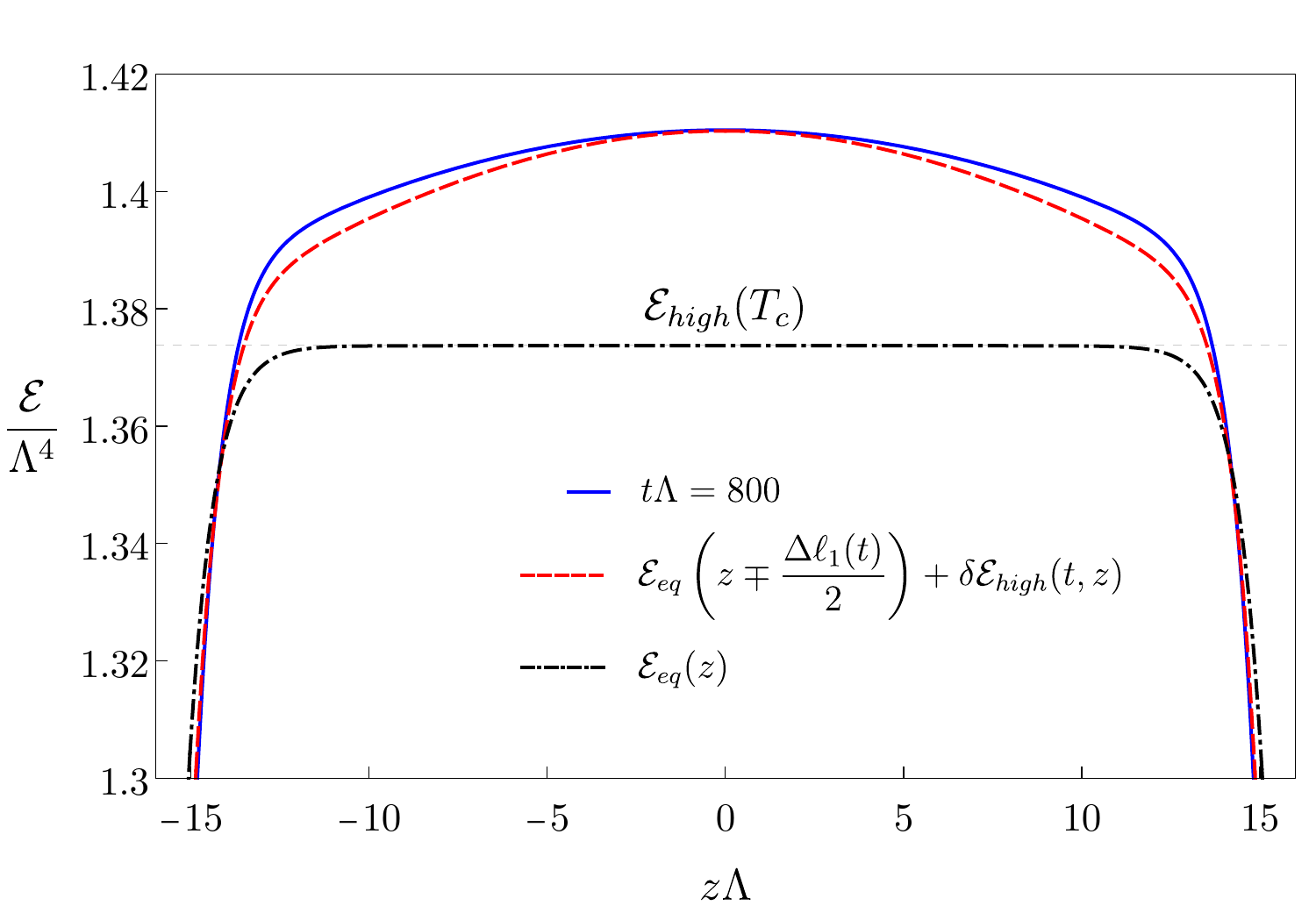} 
			\includegraphics[width=.49\textwidth]{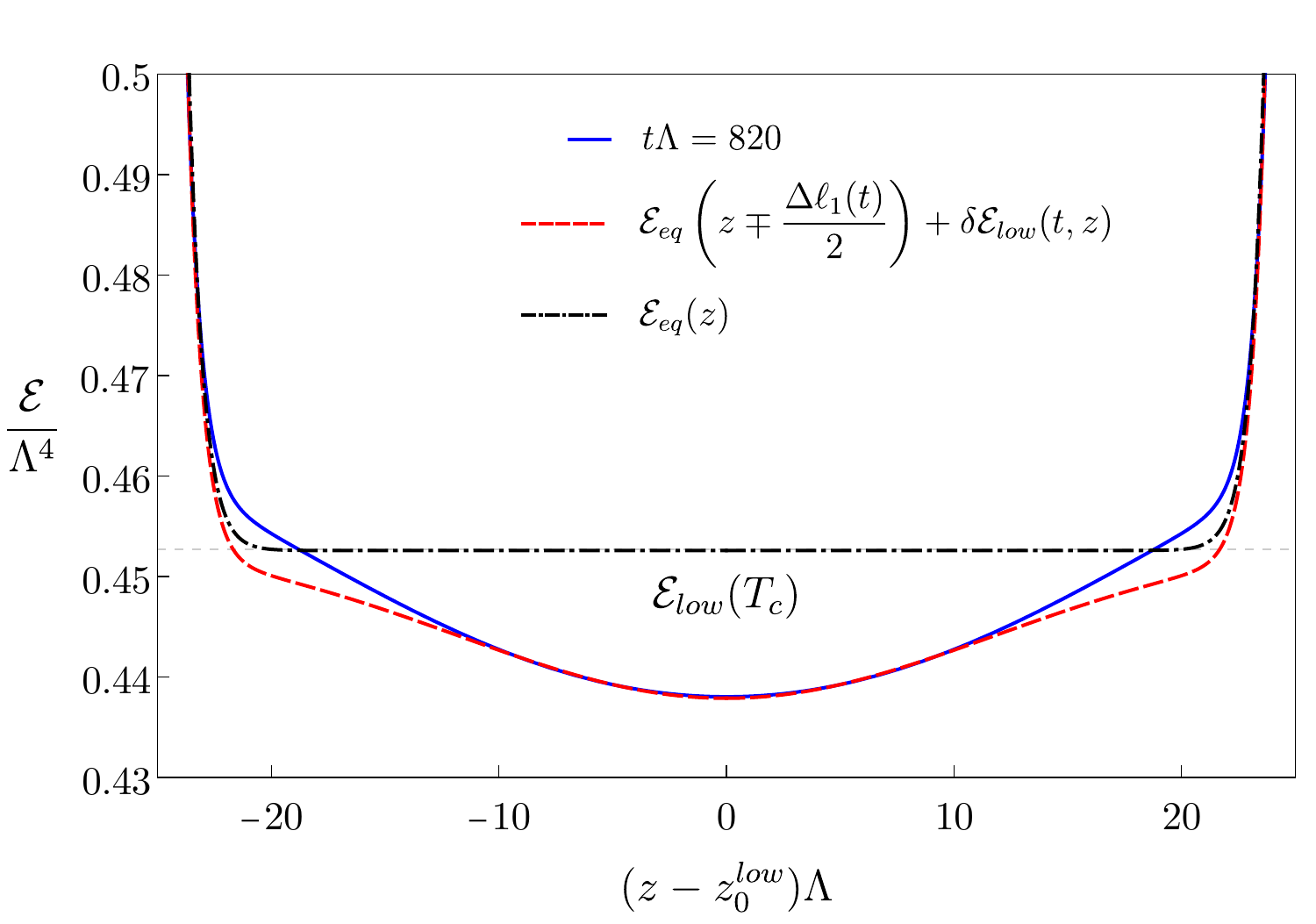} 
		\end{tabular}
	\end{center}
	\vspace{-5mm}
	\caption{\label{fig:E_oscilations_005} \small Oscillations of the high- and 
	low-energy phases during the relaxation period after the merger.}
	\vspace{7mm}
\end{figure}

The collision takes place around $t\Lambda\sim 500$. The two domains merge into a single, out-of-equilibrium, larger domain that will eventually reach equilibrium through damped oscillations. At sufficiently late times we expect to be able to describe the dynamics in terms of an equilibrium profile plus small deviations, $\delta\mathcal{E}/\mathcal{E}\ll 1$, so that we can apply a linear analysis.
Following \cite{Attems:2019yqn}, we model the system at late times as two phases of energies $\mathcal{E}_{low}$ and $\mathcal{E}_{high}$ separated by two interfaces or walls that can move as a whole but whose shapes are rigid. Since the system is symmetric under $z\to -z$ we will often focus only on $z>0$ and speak of ``the'' wall. We define the positions of the wall as the point at which the energy density is the average between $\mathcal{E}_{high}$ and $\mathcal{E}_{low}$, namely we require that $\mathcal{E}(t, z_{wall}) = (\mathcal{E}_{high}+\mathcal{E}_{low})/2$. The size of the high-energy phase is then $\ell_1 = 2 z_{wall}$ and the size of the low-energy phase is $\ell_2=L-\ell_1$.  Rigidity implies that the perturbations of each phase  must vanish at the wall, meaning that the only allowed perturbations are odd cosine harmonics of wavelength $\lambda_i=2\ell_i$. In general there could also be sine harmonic of wavelength $\lambda_i=\ell_i$, but these are excluded in our particular case because of the reflection symmetry. As both subsystems are coupled and the wall is rigid we expect the two phases to oscillate in time as a whole, with identical frequencies. Since translation invariance is broken, all the modes in the Fourier decomposition of $\delta \mathcal{E}=\mathcal{E}(t,z)-\mathcal{E}_{eq}(z)$, with $\mathcal{E}_{eq}(z)$ the equilibrium profile at asymptotic times,  will mix with one another. Thus, at late times they should all oscillate with the frequency of the longest-lived mode (see Fig.~25 in \cite{Bea:2020ees} for a related discussion). We verified this for modes up to $n=14$, i.e.~for modes with momentum up to $k=14 \times 2\pi/L$. \fig{fig:Fourier_005} illustrates the result for the four lowest  modes. 
This result motivates the following ansatz for the perturbations of each phase at late times:
\begin{equation}
\delta \mathcal{E}_i(t,z)=
\begin{cases}
e^{\omega_1(t-t_0)}\sum_{n=0}^{\infty}a_n^i\cos\Big[\omega_2(t-t_0)+\gamma_n^i\Big]\cos\Big[ \frac{(2n+1)\pi}{\ell_i}\left(z-z_0^i\right)\Big],\quad \left\vert z-z_0^i\right\vert \leq \frac{\ell_i}{2}\\
0,\quad\mathrm{otherwise},
\end{cases}
\label{eq:perturbation_oscillations}
\end{equation}
where the index $i$ refers to each of the two phases and $z_0^i$ refers to their midpoints. The form of the energy density at late times is then
\begin{equation}
\mathcal{E}(t,z)=\mathcal{E}_{eq}\left(z\mp\frac{\Delta \ell_1(t)}{2}\right)+
\delta \mathcal{E}_{high}(t,z) + \delta \mathcal{E}_{low}(t,z),
\label{eq:fit_oscillations}
\end{equation}
where $\mp$ means that we use $-$ ($+$) if $z> 0$ ($z<0$). The second and third terms on the right-hand side describe the oscillation in time and space of the two phases. The first term is due to the rigid motion of the walls as the phase domains oscillate. If we call $\Delta \ell_1(t)=-\Delta \ell_2 (t)$ the changes in the sizes of the two phases caused by the motion of the walls,  then mathematically this means that the  oscillations happen on top of phases with sizes $\ell_i+ \Delta \ell_i(t)$, whose energy density can be written approximately as the first term in (\ref{eq:fit_oscillations}). The form of this term is just a simple way of ``stretching'' (for positive $\Delta \ell_i$) or ``compressing'' (for negative $\Delta \ell_i$) the domain profile by ``gluing in''  or ``cutting out'' a small piece at the centre of each phase, taking advantage of the fact that the energy density is almost exactly constant there. In order to determine $\Delta \ell_1$ as a function of time, we simply impose conservation of energy, namely that the energy change associated to the rigid shift of the walls is exactly compensated by the energy change associated to the oscillations of the phases: 
\begin{equation}
2 \left( \mathcal{E}_{high}-\mathcal{E}_{low} \right)  \Delta \ell_1(t) + \int_{-L/2}^{L/2}dz \Big[ \delta \mathcal{E}_{high}(t,z)+\delta \mathcal{E}_{low}(t,z)\Big] =0 \,.
\end{equation}
The approximation (\ref{eq:fit_oscillations}) becomes applicable soon after the merger, as illustrated by  \fig{fig:E_oscilations_005}.
To describe correctly the perturbations of the high-energy phase it is enough to truncate the series \eqref{eq:perturbation_oscillations} at $n=1$, whereas in order to obtain a good result for the low-energy phase we included the second mode, $n=2$. The values that we found for the fit parameters  in \eqref{eq:perturbation_oscillations} are
\begin{equation}
\begin{array}{c c c}
\omega_1/\Lambda = -0.0048, & \omega_2/\Lambda = 0.032, & t_0\Lambda=800, \\
L_{high}\Lambda=33.33, & L_{low}\Lambda=50.82, & L\Lambda = 84.14,\\
z_0^{high} = 0, & z_0^{low} = \pm L/2, & a_1^{high}/\Lambda^4 =0.0375, \\ \gamma_1^{high}= 0.227, & a_1^{low}/\Lambda^4 = 0.0142, & \gamma_1^{low}= 2.346,\\
a_2^{low}/\Lambda^4 = -0.0022, & \gamma_2^{low} =-0.244.
\end{array}\\
\end{equation}

\subsection{Intermediate velocity}
\begin{figure}[t!!!]
\begin{center}
	\includegraphics[width=.65\textwidth]{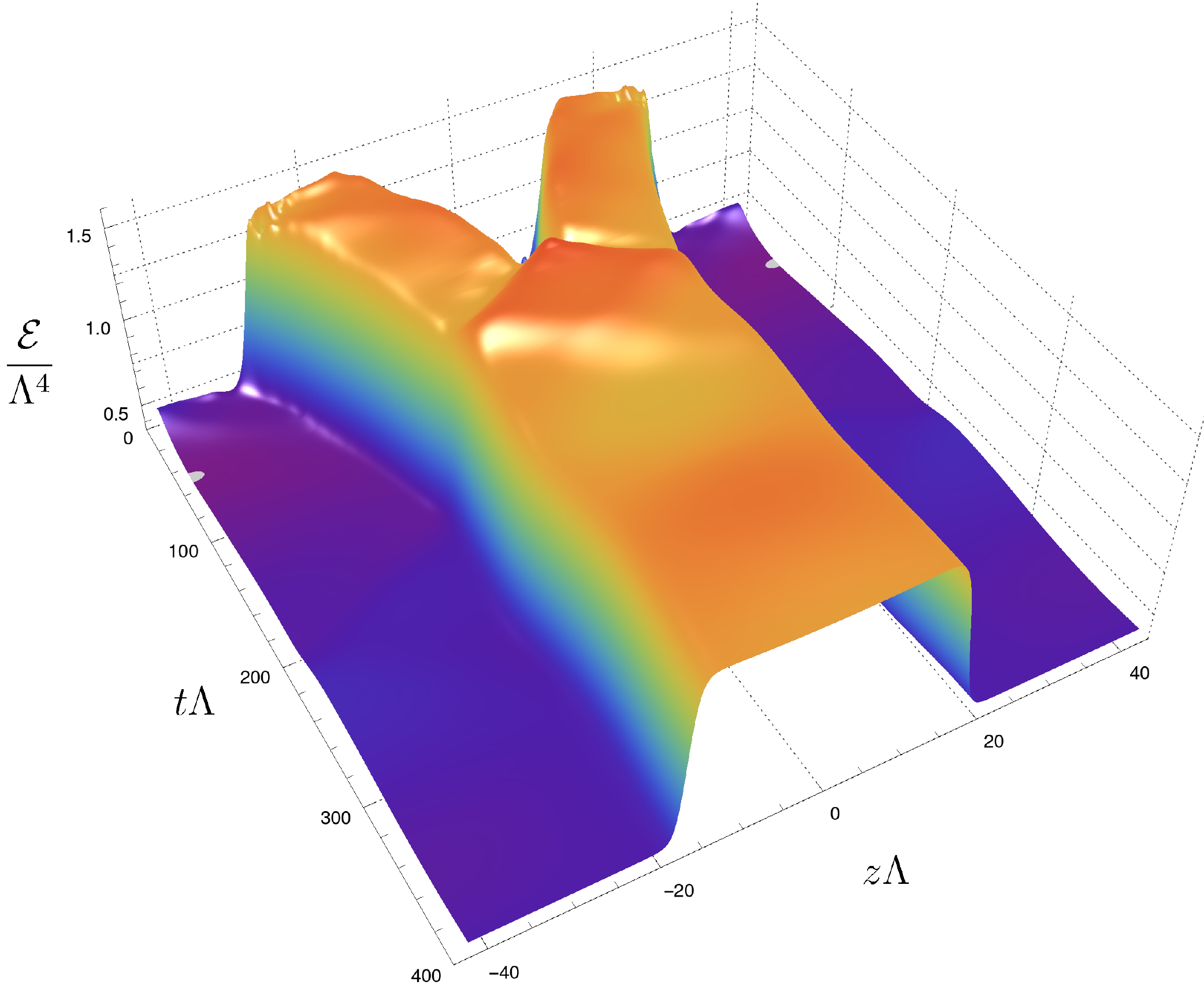} \\[2mm]
	\includegraphics[width=.65\textwidth]{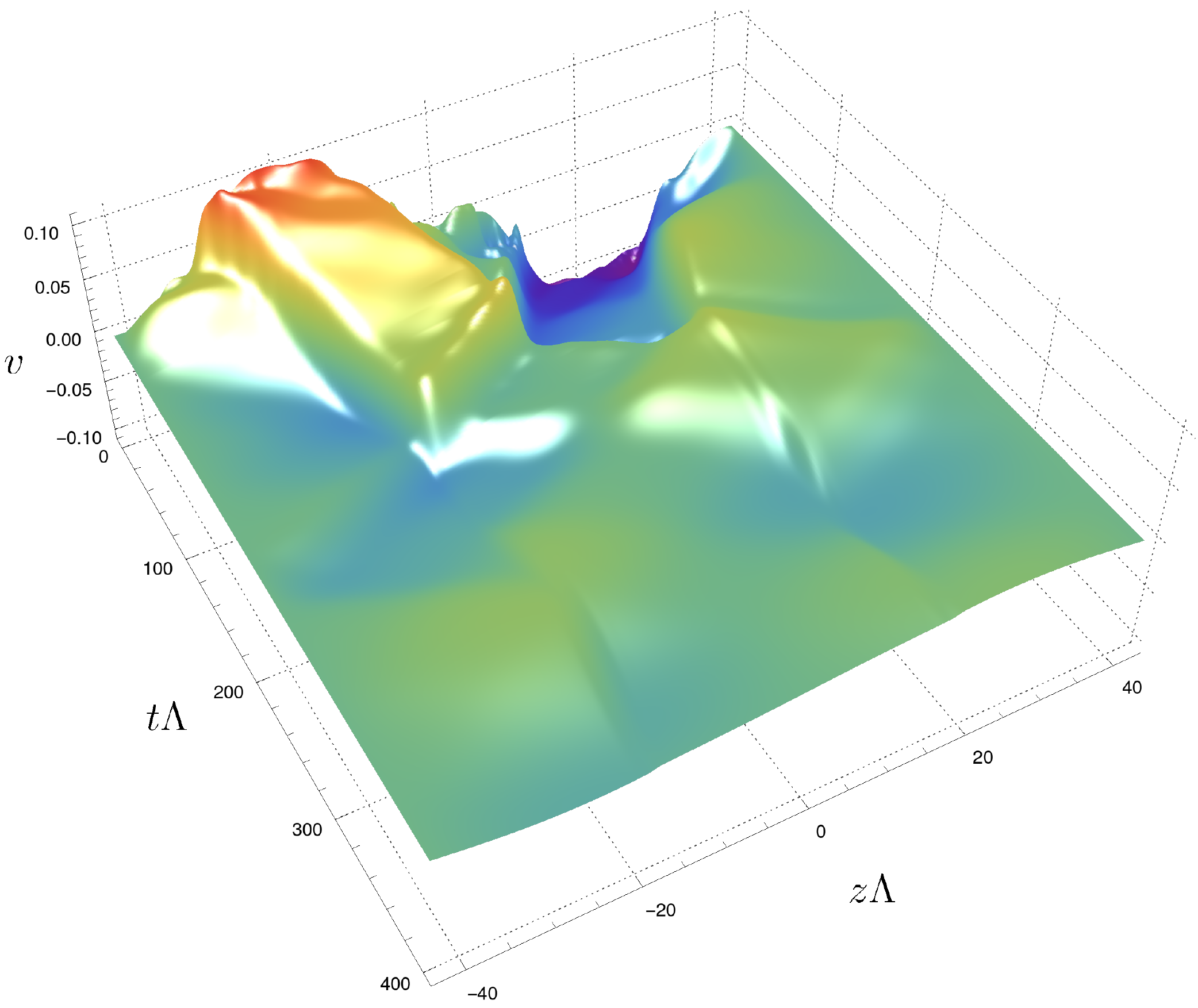} 
\end{center}
\vspace{-5mm}
\caption{\small
Evolution of the energy density (top) and of the fluid velocity (bottom) for a simulation with $v_{\mathrm{max}}= 0.11$.}
\vspace{0mm}
\label{fig:3D_vmax0092}
\end{figure}

\begin{figure}[h!!!]
\begin{center}
	\includegraphics[width=.6\textwidth]{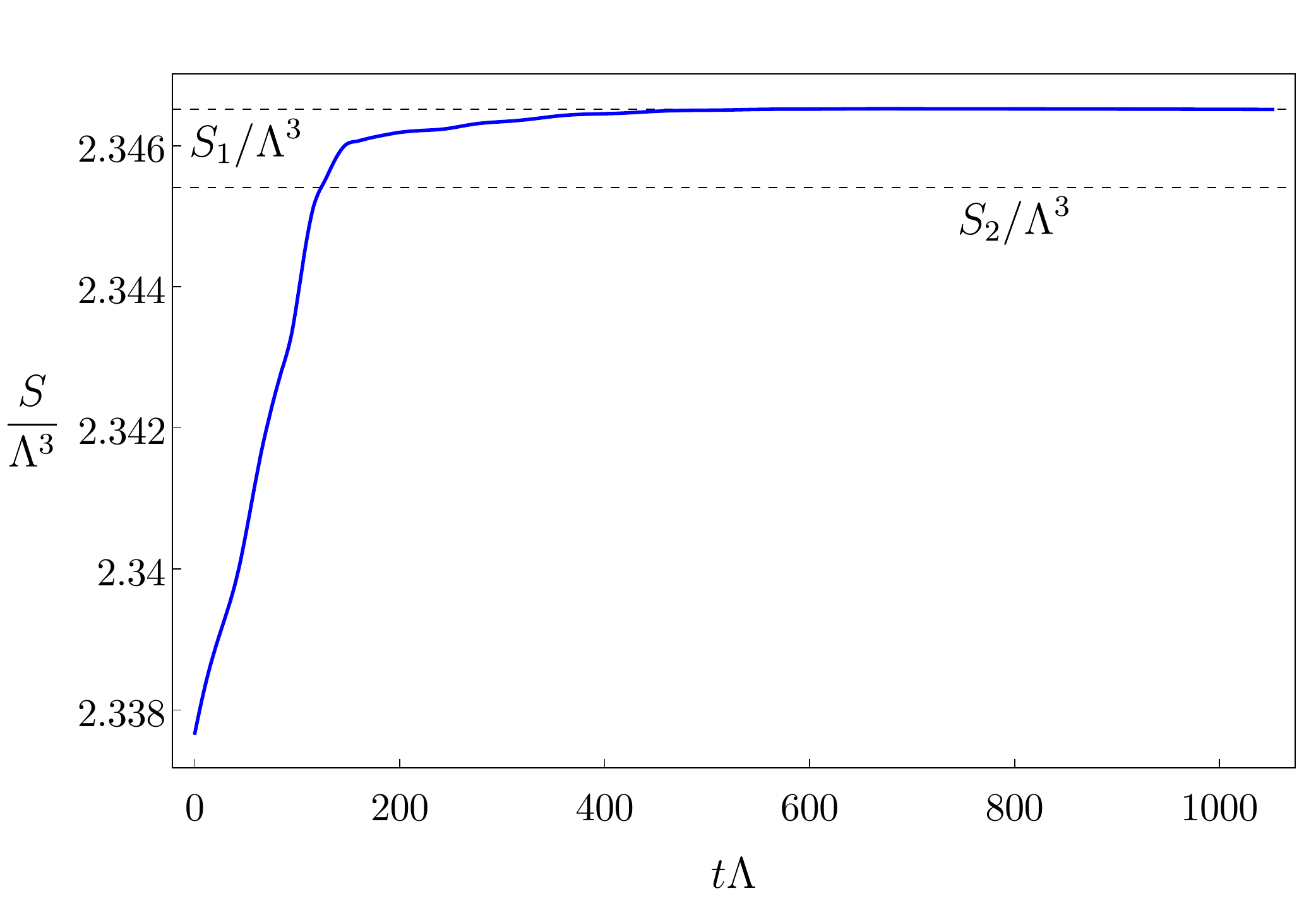} 
\end{center}
\vspace{-5mm}
\caption{\label{fig:vmax_0091_S} \small Time evolution of the total entropy (per unit transverse area) for the evolution in \fig{fig:3D_vmax0092}. $S_2$ is the entropy of a static configuration with two domains (the dashed, red curve in \fig{fig:merging_initial}). $S_1$ is the entropy of a static configuration with the same total energy but a single domain.}
	\vspace{2mm}
\end{figure}

A simulation in this regime with $v_{max}=0.11$ is shown in \fig{fig:3D_vmax0092}. 
In this case there is no quasi-static regime, as seen by the absence of an intermediate plateau in the entropy density in \fig{fig:vmax_0091_S}. The overall increase in the entropy is  greater than in the low-velocity case, confirming the intuition that this is a more violent collision. 
\begin{figure}[h!!!]
	\begin{center}
		\begin{tabular}{cc}
			\includegraphics[width=.45\textwidth]{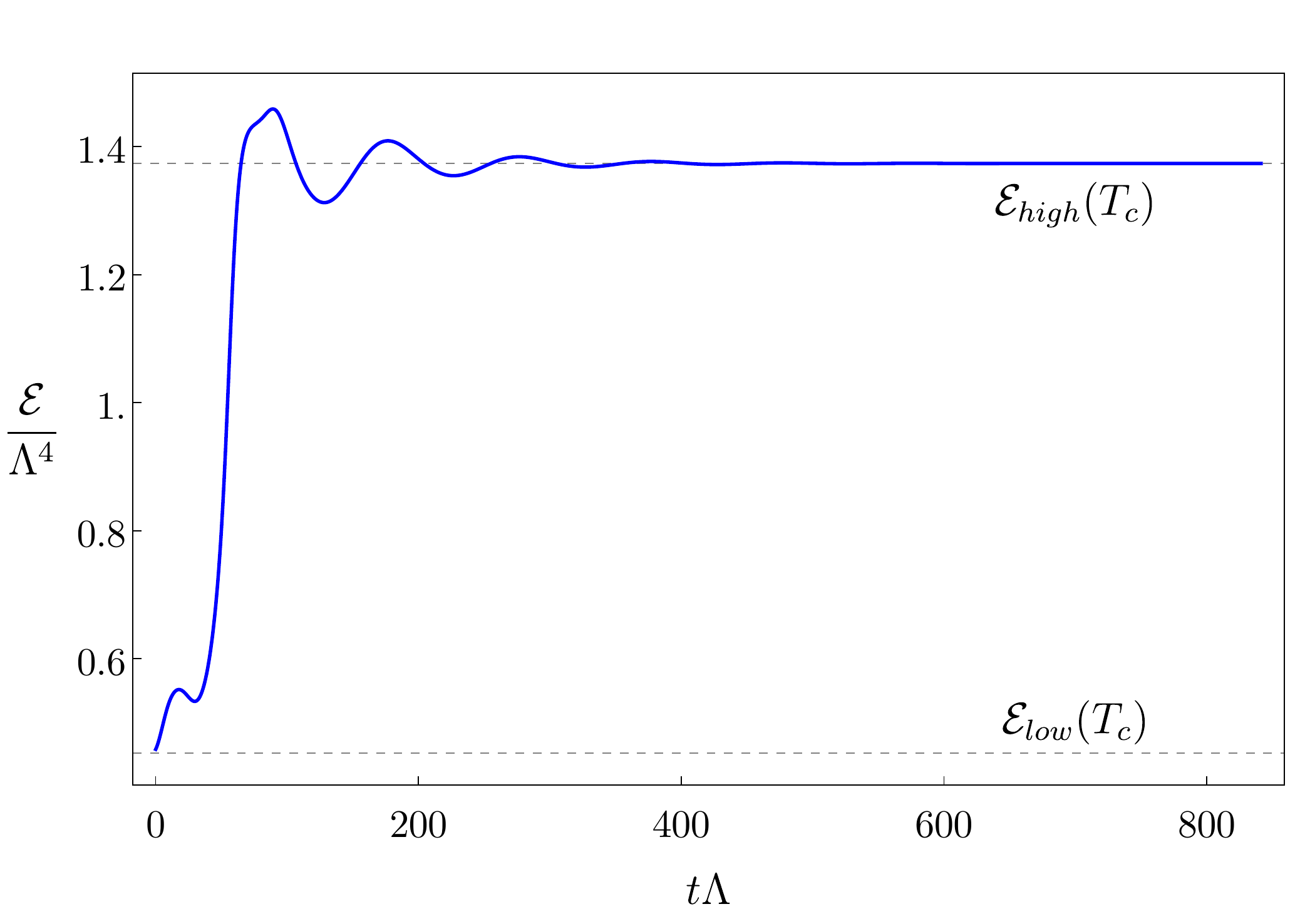} 
			\includegraphics[width=.45\textwidth]{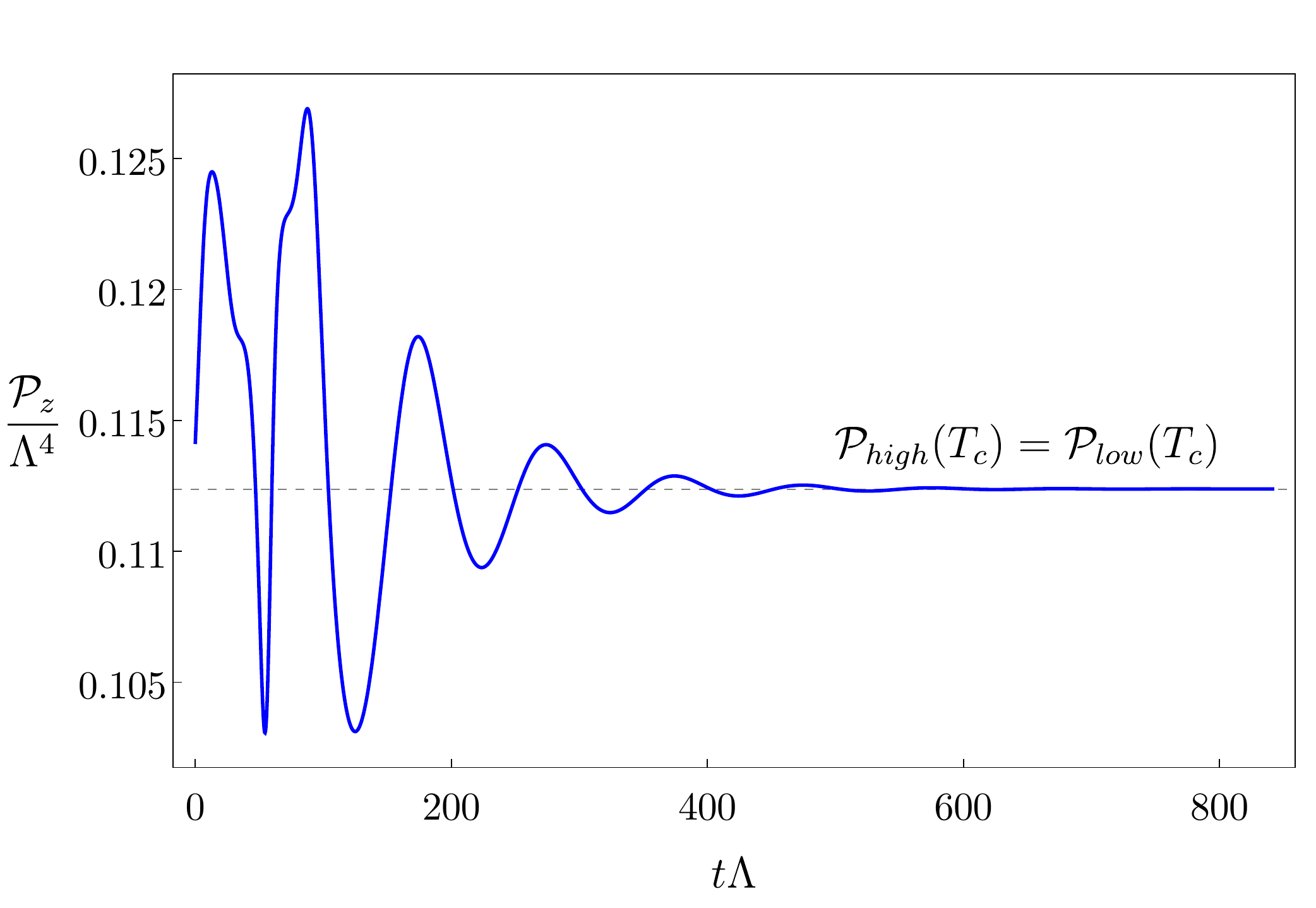} 
		\end{tabular}
	\end{center}
	\vspace{-5mm}
	\caption{\label{fig:E_P_middle_low_0091} \small (Left) Energy density $\mathcal{E}/\Lambda^4$ and (Right) longitudinal pressure $P_l/\Lambda^4$ as a function of time for constant coordinate $z\Lambda=0$ for the evolution in \fig{fig:3D_vmax0092}.}
	\vspace{2mm}
\end{figure}
The rest of the evolution is qualitatively similar to the low-velocity case: there is an initial excess of pressure between the domains that slows them down and they  merge to form a domain that relaxes to equilibrium through the same kind of oscillations, see \fig{fig:E_P_middle_low_0091}.
\begin{figure}[h!!!]
	\begin{center}
		\begin{tabular}{cc}
			\includegraphics[width=.5\textwidth]{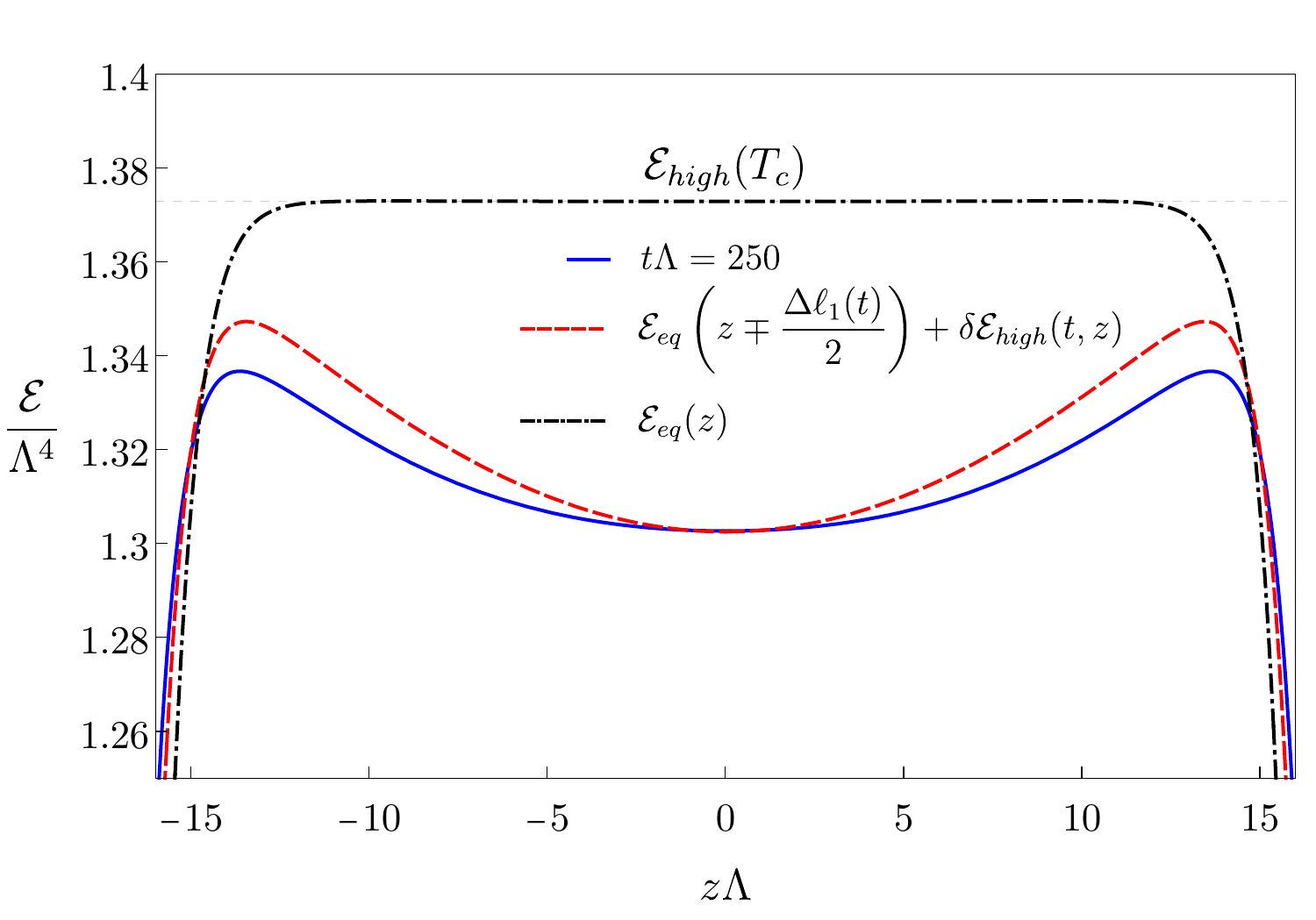} 
			\includegraphics[width=.5\textwidth]{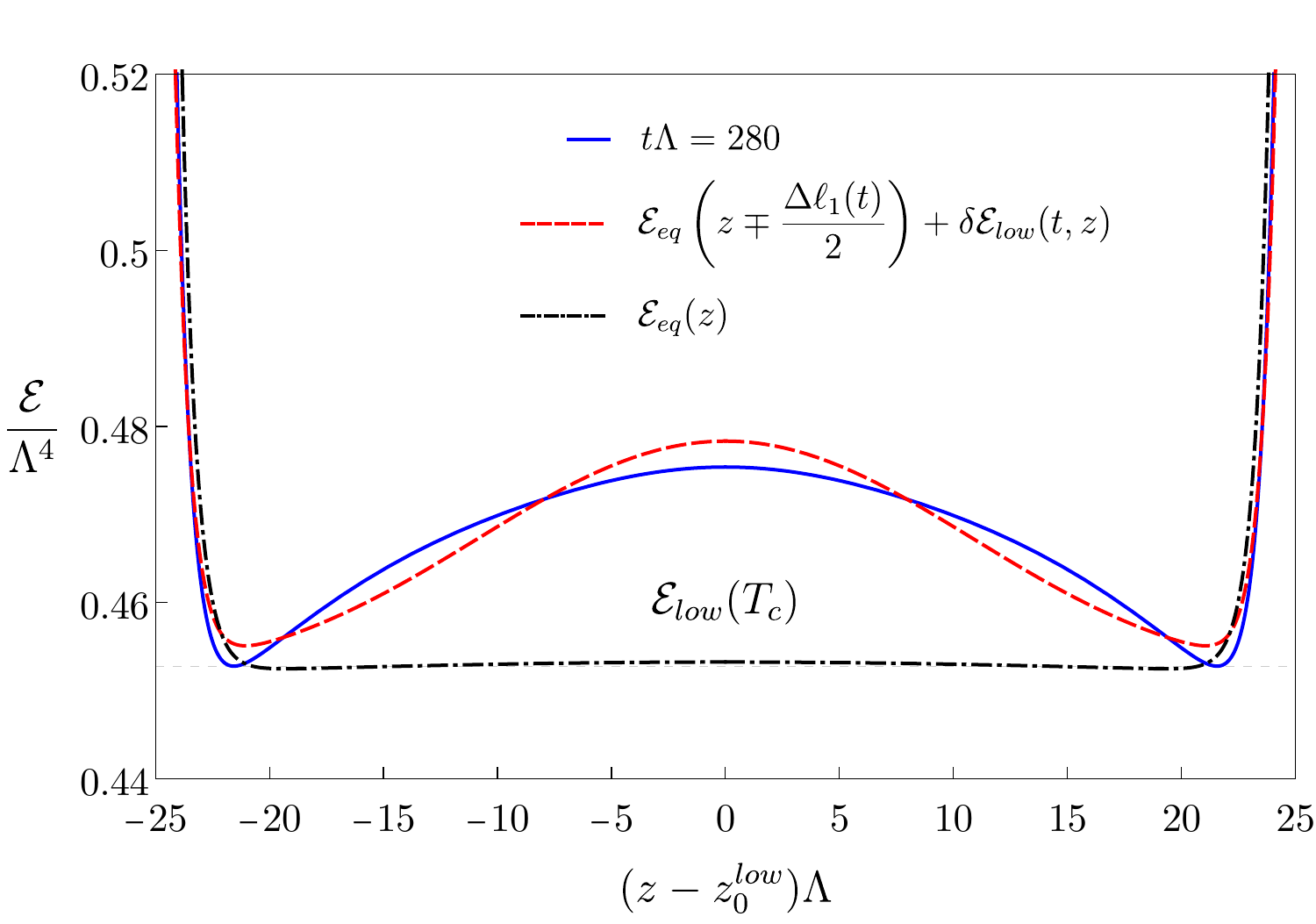} 
		\end{tabular}
	\end{center}
	\vspace{-5mm}
	\caption{\label{fig:E_oscilations_0091} \small Oscillations of the high- and 
	low-energy phases during the relaxation period after the merger for the evolution in \fig{fig:3D_vmax0092}.}
	\vspace{0mm}
\end{figure}
Applying \eqref{eq:fit_oscillations} leads to a good description  of the oscillations at  times soon after the collision, see \fig{fig:E_oscilations_0091}. The parameters in this case are
\begin{equation}
\begin{array}{c c c}
\omega_1/\Lambda = -0.0047, & \omega_2/\Lambda = 0.032, & t_0\Lambda=200, \\
L_{high}\Lambda=33.33, & L_{low}\Lambda=50.82, & L\Lambda = 84.14,\\
z_0^{high} = 0, & z_0^{low} = \pm L/2, & a_1^{high}/\Lambda^4 =0.1065, \\ \gamma_1^{high}= 2.09, & a_1^{low}/\Lambda^4 = -0.0397, & \gamma_1^{low}= 1.306,\\
a_2^{low}/\Lambda^4 = -0.0082, & \gamma_2^{low} =1.26\,.
\end{array}\\
\end{equation}
Note that, up to fitting errors, the frequencies in this case and in the low-velocity case are the same. The reason is that the final equilibrium state is the same in both cases, and therefore the frequencies of the linear perturbations around it are identical. The only difference between the simulations is therefore in the amplitudes of the oscillations.
%
%
The following table collects the ratios of the coefficients for the $n=1, 2$ modes, $a_2^i/a_1^i$, for both phases for various simulations:
\begin{equation}
\begin{array}{|c| c| c|}
\hline
v_{max} & \mathrm{low} & \mathrm{intermediate}\\
\hline
\mathrm{low-energy \,\, phase} & -0.158 & 0.2058 \\
\hline
\mathrm{high-energy \,\, phase} & -0.043 & -0.044\\
\hline
\end{array}
\end{equation}
We see that they are of the same order for different simulations, and that the ratio is smaller in the high-energy phase, consistently with our neglect of the $n=2$ (and higher) modes in this phase.  

\subsection{High velocity}
\begin{figure}[h!!!]
	\begin{center}
			\includegraphics[width=.8\textwidth]{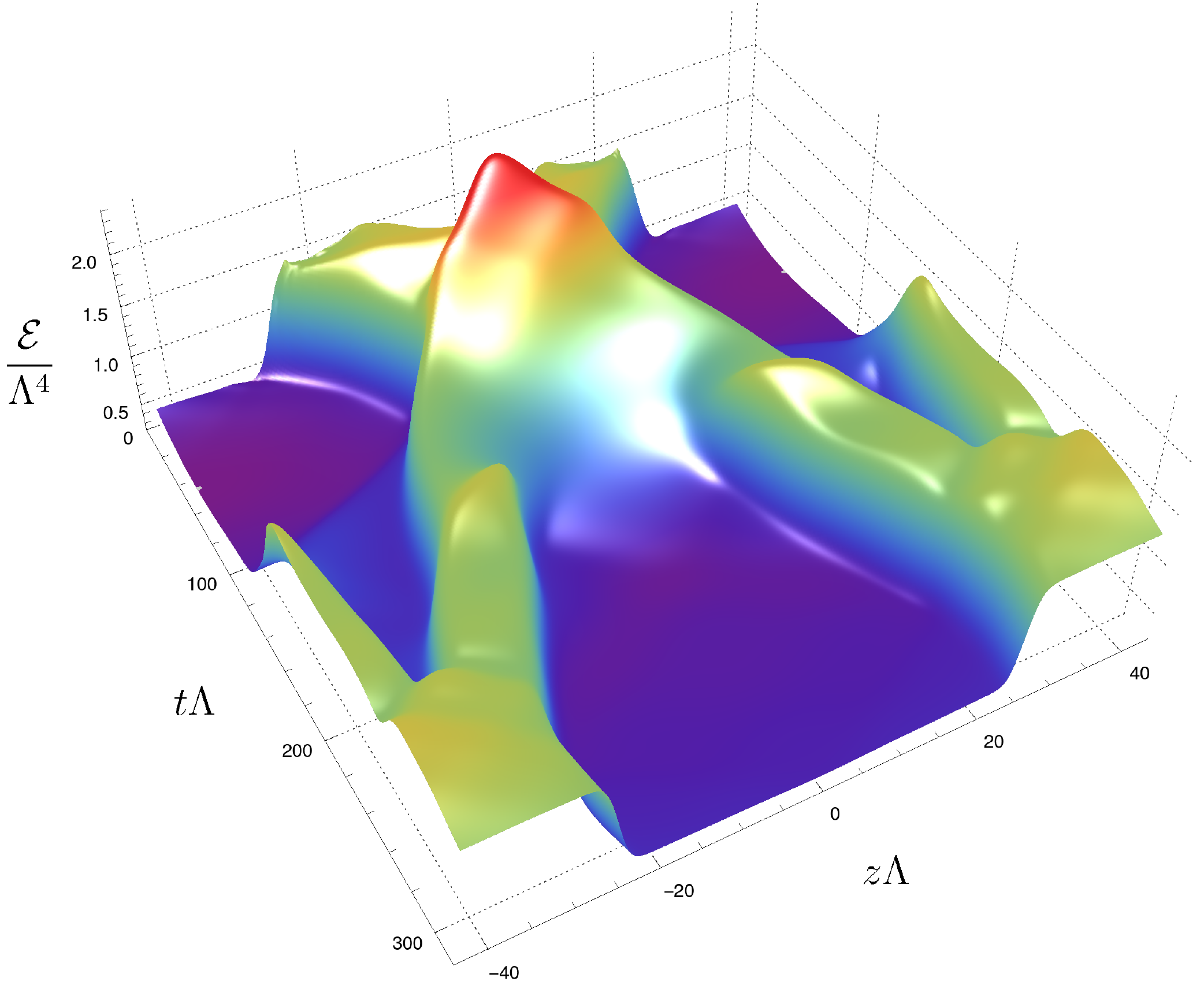} \\[2mm]
			\includegraphics[width=.8\textwidth]{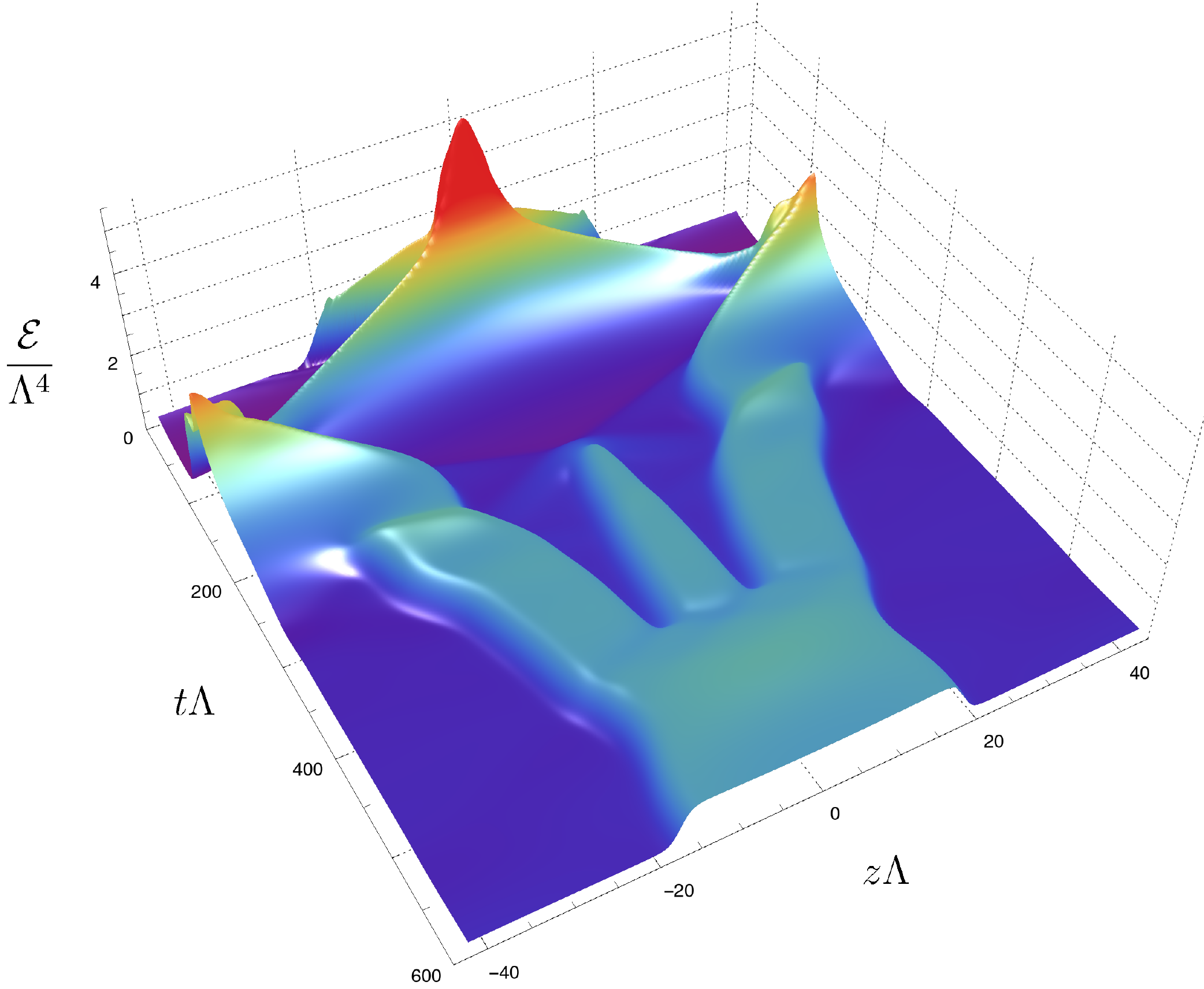}  
	\end{center}
	\vspace{-2mm}
	\caption{\label{fig:3D_high_speed} \small Evolution of the energy density for  simulations with $v_{\mathrm{max}}= 0.25$ (top) and $v_{\mathrm{max}}= 0.73$ (bottom).}
	\vspace{0mm}
\end{figure}

\begin{figure}[t]
\begin{center}
	\includegraphics[width=.66\textwidth]{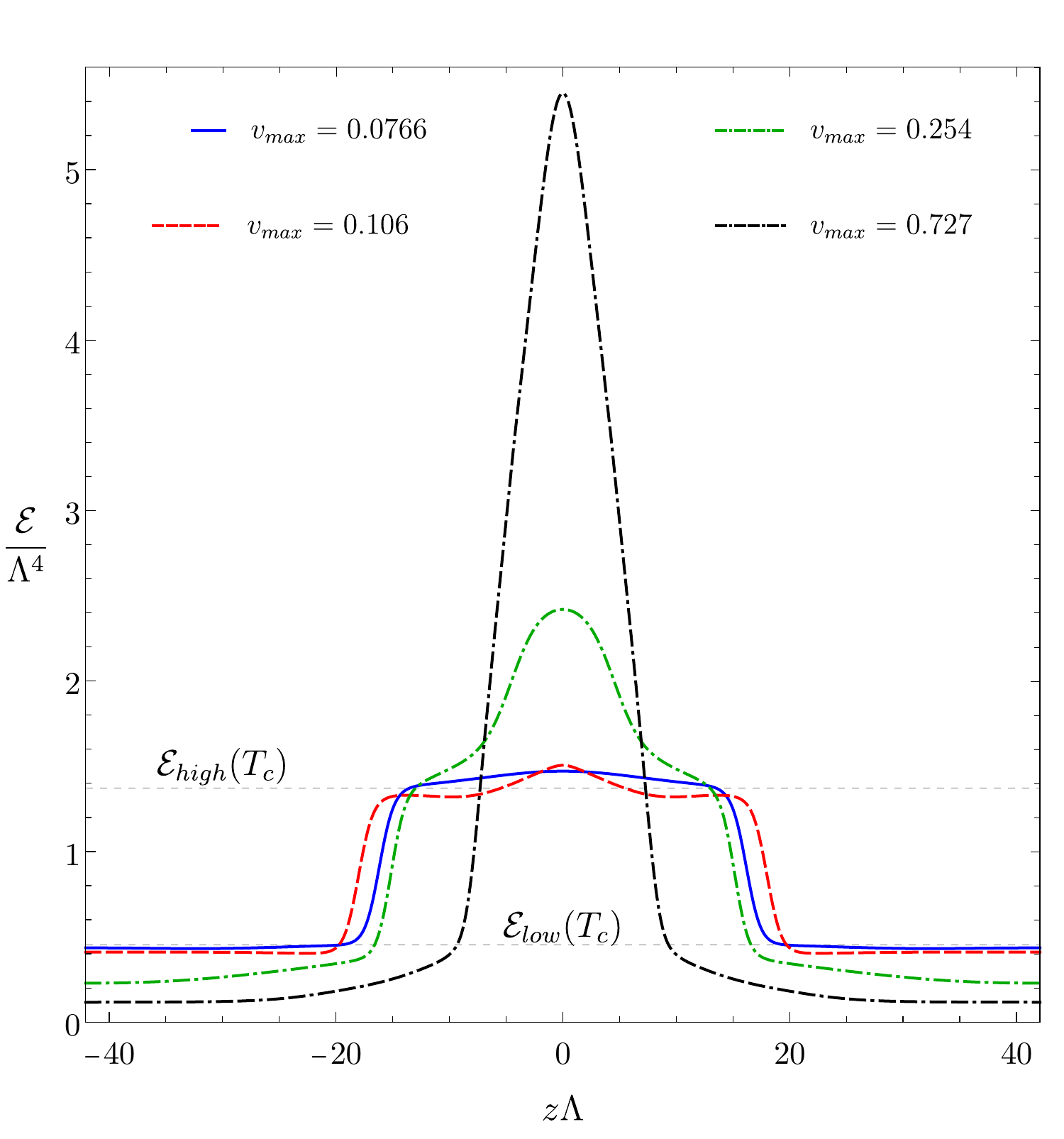} 
\end{center}
\vspace{-5mm}
\caption{\label{fig:blob_comparison} \small Snapshots of the energy density at the times at which the maximum energy is reached for collisions with different $v_{max}$.}
	\vspace{0mm}
\end{figure}

For velocities $v_{max} \geq v_2 = 0.2$ the domains collide but do not merge in the first collision. Instead, the excited state that results from the collision breaks apart into pieces that will subsequently collide again until they finally merge.  \fig{fig:3D_high_speed} shows two illustrative cases. 
As shown by \fig{fig:blob_comparison}, the peak energy density right after the first collision increases significantly with $v_{max}$, meaning that the resulting configuration is a large deviation from an equilibrium domain. The maximum relative deformations are $\mathcal{E}_{peak}/\mathcal{E}_{high}\simeq 1.071, 1.096, 1.762, 3.966$, respectively, for the increasing velocities shown in \fig{fig:blob_comparison}. As a consequence, non-linear dynamics becomes important and the excited state breaks apart into smaller components rather than relaxing to a single equilibrium domain as in previous sections. 

\begin{figure}[h!!!]
	\begin{center}
			\includegraphics[width=.60\textwidth]{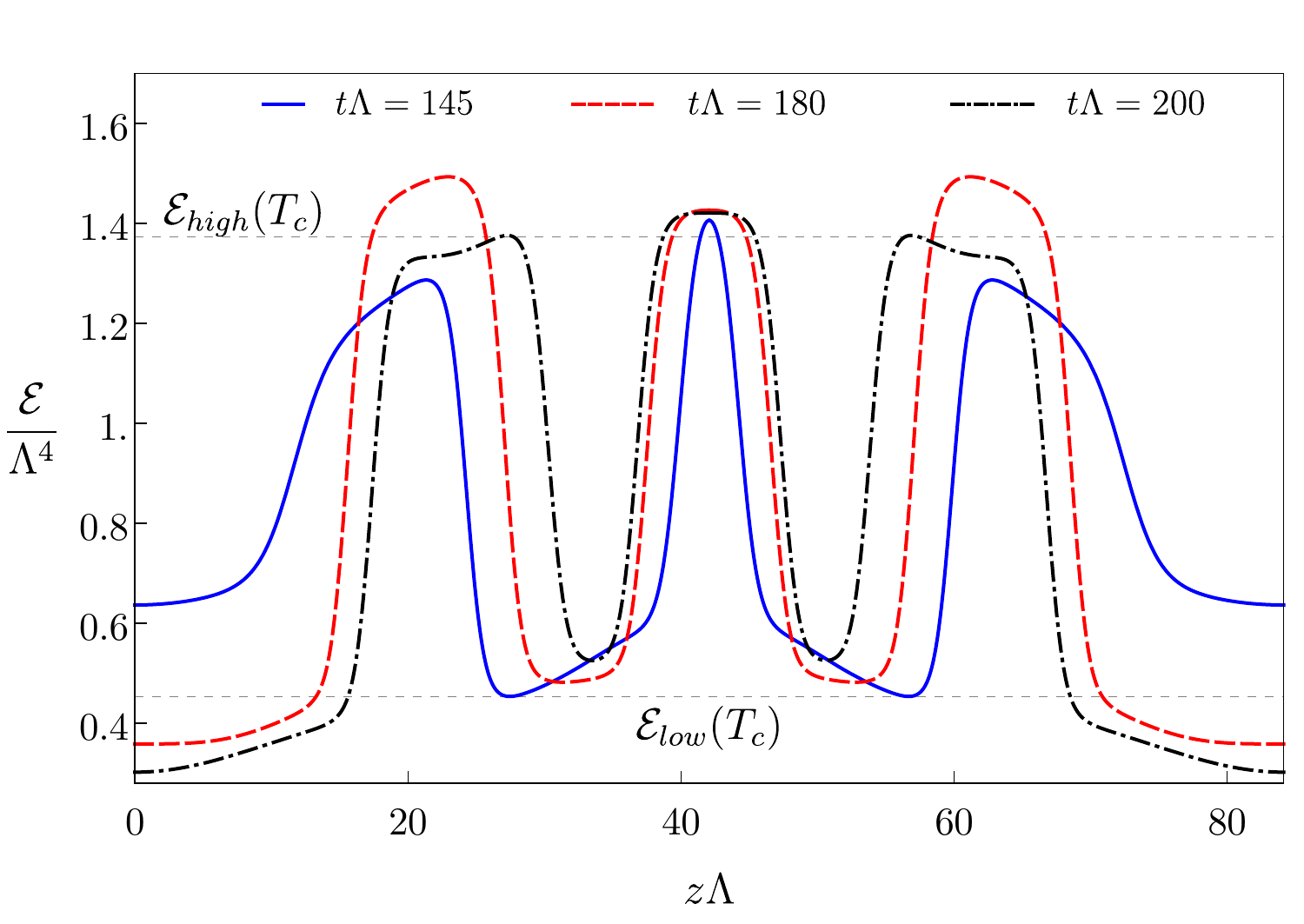} 
	\end{center}
	\vspace{-5mm}
	\caption{\label{fig:fragments_peak_025} \small Snapshots of the energy density for the collision of \fig{fig:3D_high_speed}(top) with $v_{max}=0.25$. The blue curve corresponds to a time soon after the fragmentation of the excited state. Note that the horizontal axis  in this figure has been shifted to show the new domain at the center of the figure and the central fragments on the sides of it.}
	\vspace{0mm}
\end{figure}
\begin{figure}[t]
	\begin{center}
			\includegraphics[width=.60\textwidth]{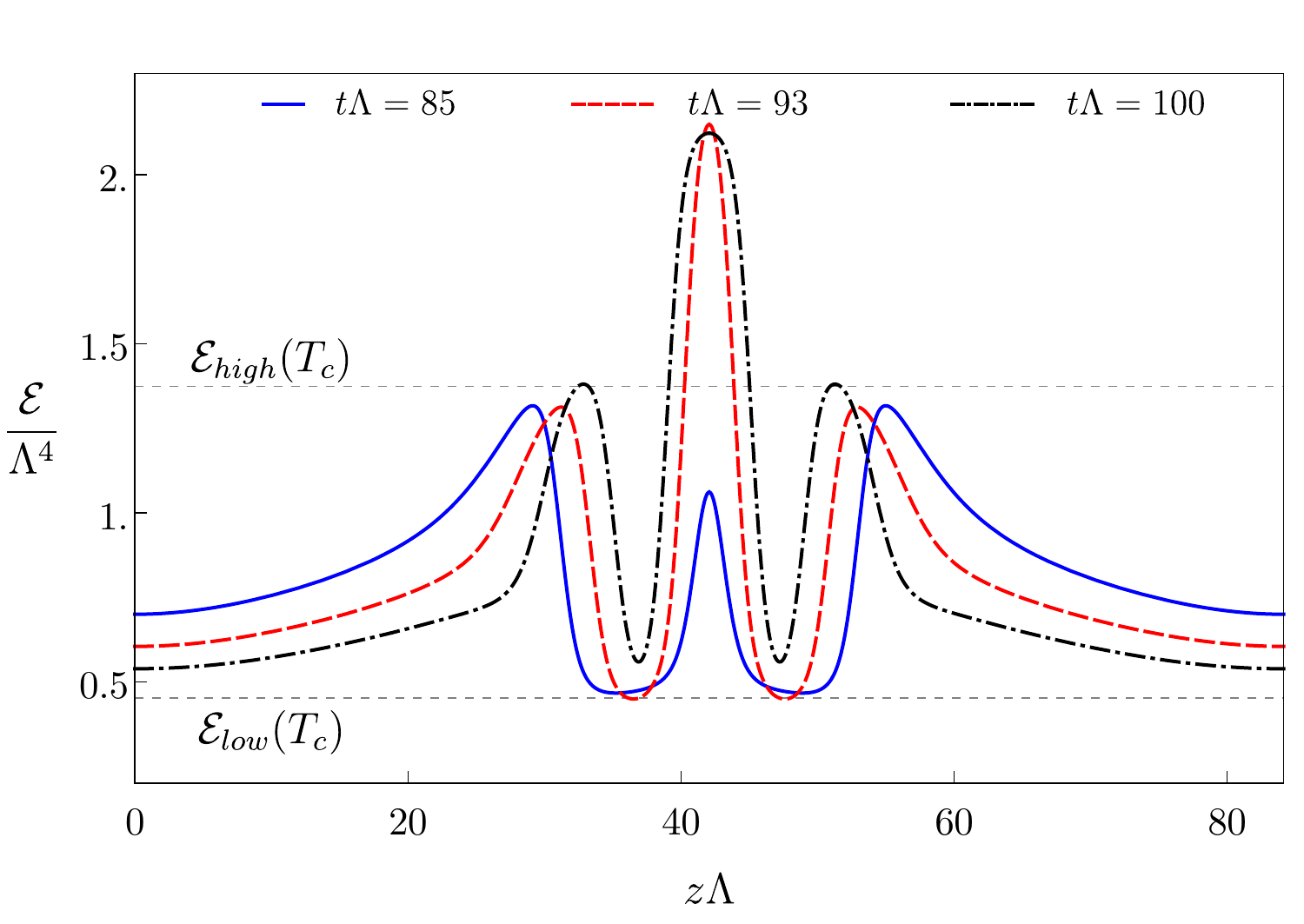} \\[2mm]
			\includegraphics[width=.60\textwidth]{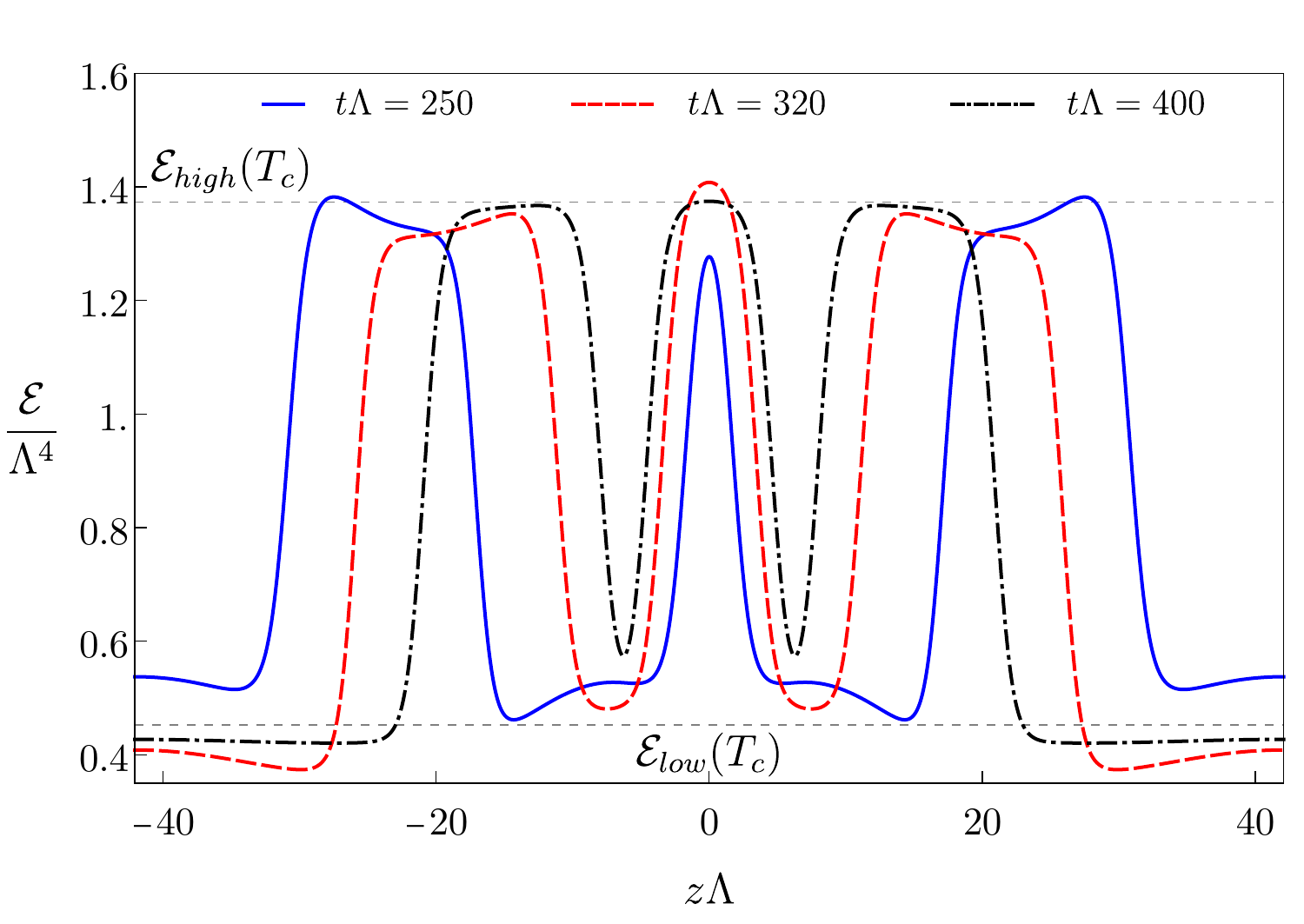} 
	\end{center}
	\vspace{-5mm}
	\caption{\label{fig:fragments_peak_073} \small 
	Snapshots of the energy density for the evolution in \fig{fig:3D_high_speed}(bottom) with $v_{max}=0.73$. The top (bottom) plot corresponds to a time shortly after the second (fourth) collision. The blue curve corresponds to a time soon after the fragmentation of the excited state.}
	\vspace{0mm}
\end{figure}

In both cases shown in \fig{fig:3D_high_speed},  the initial excited state first ``emits'' perturbations that travel away from the central blob, and subsequently it splits into two fragments. The perturbations travel away from the central blob, colliding with each other on the other side of the box, at $z\Lambda\simeq 42$. We will refer to this as the second collision. This collision creates an excitation that quickly becomes a new domain  smaller than the central fragments, as illustrated in  \fig{fig:fragments_peak_025}. Note from the horizontal axis in this figure that the new domain is shown in the center of the figure and the central fragments on the sides of it. At later times the two central fragments reach the location of the new domain and a third collision, this time a three-body one, takes place. In cases with $v_{max}\gtrsim v_2$, such as the one in \fig{fig:3D_high_speed}(top), this third collision results in the formation of a single domain that then relaxes to equilibrium as described in previous sections. 

In contrast, in cases with large enough $v_{max}$, such as the one of  
\fig{fig:3D_high_speed}(bottom), the excitations emitted by the initial excited state, as well as the fragments to which it decays, travel faster to the opposite side of the box and do not have time to relax to domains, as shown in \fig{fig:fragments_peak_073}(top). As above, the perturbations collide in what we call the second collision. However, in this case the third collision, i.e.~the collision between the resulting new structure and the fragments from the first excited state, is again a high-velocity collision. As a consequence, it results in the emission of new perturbations plus fragments that in this case travel from $z\Lambda\simeq 42$ towards $z\Lambda=0$. These products now have time to relax to approximate domains, as illustrated in \fig{fig:fragments_peak_073}(bottom). The perturbations merge in a fourth collision at $z\Lambda=0$, and finally a fifth collision occurs, again a three-body collision, which results in the formation of a single domain that relaxes to equilibrium as in previous sections.



\begin{figure}[t!!!]
	\begin{center}
		\begin{tabular}{cc}
			\includegraphics[width=.49\textwidth]{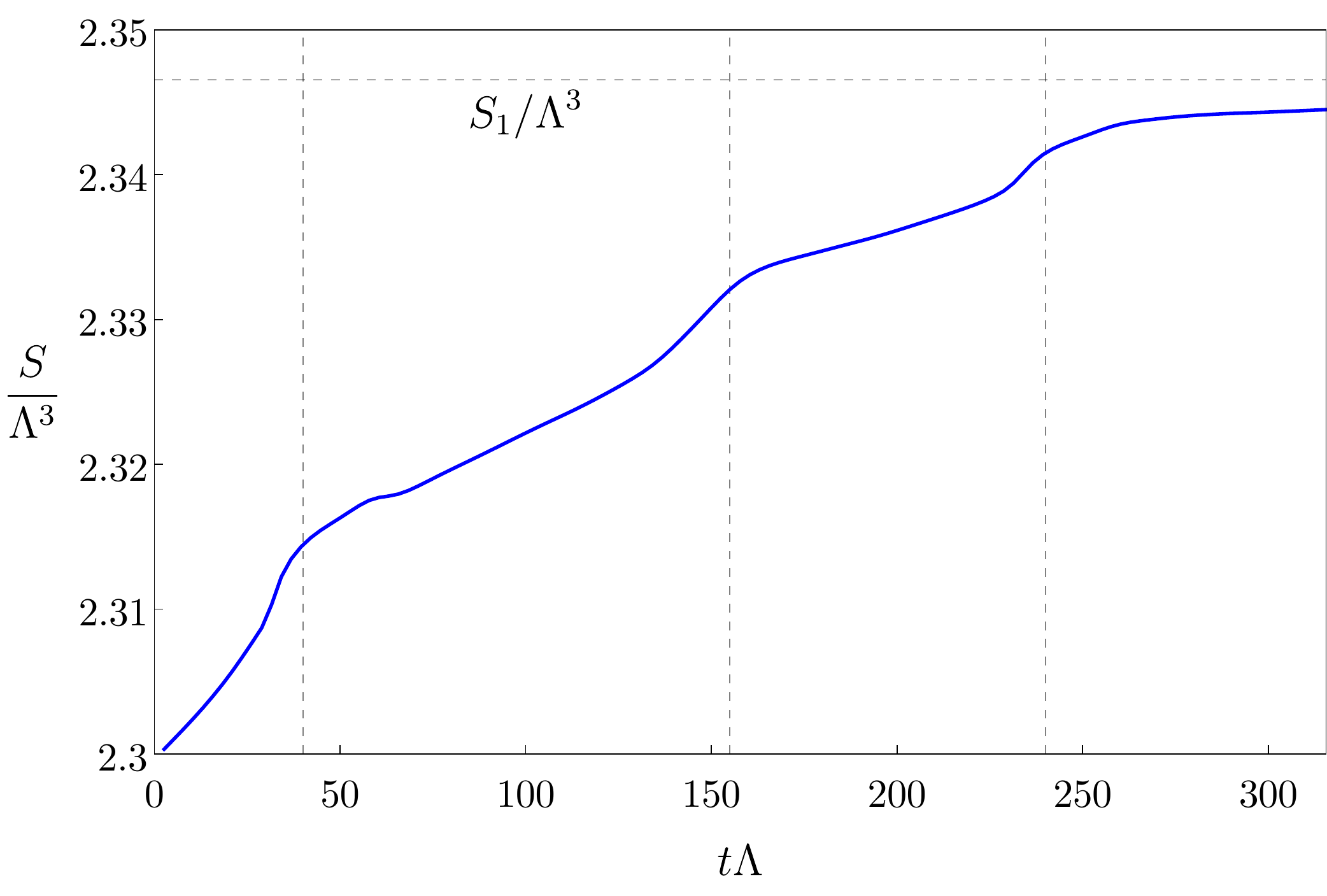} 
			\includegraphics[width=.49\textwidth]{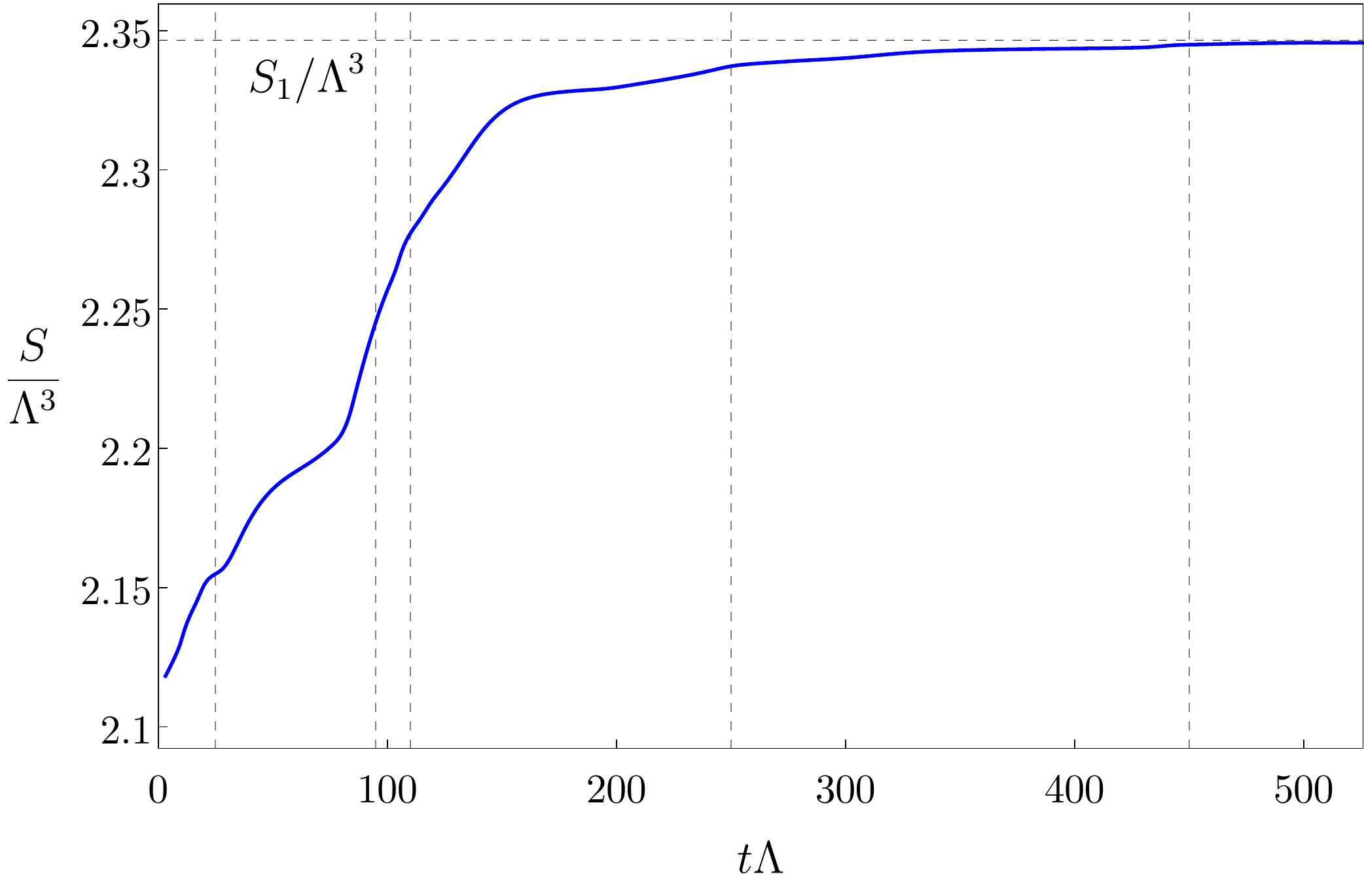} 
		\end{tabular}
	\end{center}
	\vspace{-5mm}
	\caption{\label{fig:S_high_speeds} \small Time evolution of the total entropy (per unit transverse area) for the evolutions of \fig{fig:3D_high_speed} with $v_{max}=0.25$ (left) and $v_{max}=0.73$ (right). $S_1$ is the entropy of a static configuration with the same total energy but a single domain. The vertical dashed lines represent the approximate time of each collision.} 
	\vspace{7mm}
\end{figure}

The entropy production in \fig{fig:S_high_speeds} reflects the qualitative differences described above between the two evolutions of \fig{fig:3D_high_speed}. On the left plot we can identify three different regimes corresponding to the three different collisions. In contrast, no such clear distinction is apparent on the right plot. In both cases, however, the relative amount of entropy increase is much larger than in the low- and intermediate-velocity collisions. For comparison, the relative entropy production in the evolutions with 
$v_{max}=0.11$, $v_{max}=0.25$ and $v_{max}=0.73$ are 2, 9 and 45 times larger than in the $v_{max}=0.08$ case, respectively. 


\section{Discussion}
The time evolution of the spinodal instability in a theory with a first-order phase transition typically results in the creation of phase domains that subsequently collide with one another \cite{Attems:2019yqn}. The velocities of the different domains are a complicated function of the  perturbation of the initial, homogeneous, unstable state. This makes it difficult to perform a systematic study of the physics of the collision as a function of the domain velocities. In this paper we have overcome this difficulty by directly preparing initial states consisting of domains moving towards each other at velocities $0\leq v_{max} \leq 0.73$, where $v_{max}$ is the maximum fluid velocity in the initial state. Going to higher velocities becomes challenging due to numerical issues. Nevertheless, the above range sufficed to uncover three qualitatively different dynamical regimes. 

For low velocities the domains initially slow down, enter a period of quasi-static evolution, and finally collide and merge into a single domain that then relaxes to equilibrium through damped oscillations. The quasi-static period is clearly visible as a plateau in the entropy of the system, which we computed from the area of the dual horizon on the gravity side. For intermediate velocities the evolution is qualitatively identical except for the fact that no quasi-static period is present. For high velocities the domains can collide several times before they eventually merge into a final, single domain that then relaxes to equilibrium.  Our investigations suggest that, while the precise values of the velocities that distinguish these three regimes depend on the size of the domains in question, the existence of these regimes is a robust property of the collision dynamics.


\section*{Acknowledgements}

YB acknowledges support from the European Research Council Grant No. ERC-2014-StG 639022-NewNGR and the Academy of Finland grant no. 333609. TG acknowledges financial support from FCT/Portugal Grant No.\ PD/BD/135425/2017 in the framework of the Doctoral Programme IDPASC-Portugal. MSG acknowledges financial support from the APIF program, fellowship APIF\_18\_19/226. JCS, DM and MSG are also supported by grants SGR-2017-754, PID2019-105614GB-C21, PID2019-105614GB-C22 and the ``Unit of Excellence MdM 2020-2023'' award to the Institute of Cosmos Sciences (CEX2019-000918-M). MZ acknowledges financial support provided by FCT/Portugal through the IF programme grant IF/00729/2015,
and CERN/FIS-PAR/0023/2019.
The authors thankfully acknowledge the computer resources, technical expertise and assistance provided by CENTRA/IST. Computations were performed in part at the cluster ``Baltasar-Sete-S\'ois'' and supported by the H2020 ERC Consolidator Grant ``Matter and strong field gravity: New frontiers in Einstein's theory'' grant agreement No.\ MaGRaTh-646597. We also thank the MareNostrum supercomputer at the BSC (activity Id FI-2021-1-0008) for significant computational resources.

\appendix

\end{document}